\documentclass[%
reprint,
superscriptaddress,
groupedaddress,
unsortedaddress,
runinaddress,
showpacs,
 amsmath,amssymb,
 aps,
pra,
floatfix,
]{revtex4-1}

\usepackage{xcolor}
\usepackage{marginnote}

\usepackage{graphicx}% Include figure files
\usepackage{dcolumn}% Align table columns on decimal point
\usepackage{bm}% bold math
\usepackage{enumerate}

\usepackage{hyperref}

\usepackage{orcidlink}

\newcommand{\figref}[1]{{Fig.\ref{#1}}}

\newcommand{\bra}[1]{{\langle\left.#1\right|}} %\bra{a} -> <a|
\newcommand{\ket}[1]{{\left|#1\right.\rangle}} %\ket{a} -> |a>

\newcommand{\operator}[1]{{\hat{#1}}} %consistent notation for QM operators
\begin{document}

\preprint{APS/123-QED}

\title{Hyperfine interaction in the Autler-Townes effect II: control of two-photon selection rules in the Morris-Shore basis}% Force line breaks with \\
\author{Arturs~Cinins\,\orcidlink{0000-0002-3660-2033}$^{1}$}
\email[]{arturs.cinins@lu.lv}
\author{Dmitry~K.~Efimov\,\orcidlink{0000-0002-1264-4044}$^{2}$}
\author{Martins~Bruvelis\,\orcidlink{0000-0002-1515-6820}$^{3}$}
\author{Kaspars~Miculis\,\orcidlink{0009-0009-5635-9817}$^{1,4}$}
\author{Teodora~Kirova\,\orcidlink{0000-0002-4035-1310}$^{1}$}
\author{Nikolai~N.~Bezuglov\,\orcidlink{0000-0003-0191-988X}$^{5,6}$}
\author{Igor~I.~Ryabtsev\,\orcidlink{0000-0002-5410-2155}$^{6,7}$}
\author{Marcis~Auzinsh\,\orcidlink{0000-0002-4904-5766}$^{2}$}
\author{Aigars~Ekers\,\orcidlink{0000-0001-9141-7597}$^{3}$}

\affiliation{
$^{1}$Institute of Atomic Physics and Spectroscopy, University of Latvia, Jelgavas Street 3, LV-1004 Riga, Latvia\\
$^{2}$Laser Centre, University of Latvia, Jelgavas Street 3, LV-1004 Riga, Latvia\\
$^{3}$King Abdullah University of Science and Technology (KAUST), Computer, Electrical and Mathematical Sciences and Engineering Division (CEMSE), Thuwal 23955-6900, Saudi Arabia\\
$^{4}$National Research Nuclear University MEPhI, Moscow, 115409 Russia\\
$^{5}$Saint-Petersburg State University, 199034 St. Petersburg, Russia\\
$^{6}$Rzhanov Institute of Semiconductor Physics SB RAS, 630090 Novosibirsk, Russia\\
$^{7}$Novosibirsk State University, Department of Physics, 630090 Novosibirsk, Russia\\
% $^{9}$Dipartimento di Fisica Enrico Fermi, Universit\`a di Pisa, I-56127 Pisa, Italy
}

\date{\today}% It is always \today, today,
 % but any date may be explicitly specified

\begin{abstract}
We investigated the absence of certain bright peaks in Autler-Townes laser excitation spectra of alkali metal atoms. 
Our research revealed that these dips in the spectra are caused by a specific architecture of adiabatic (or “laser-dressed”) states in hyperfine (HF) components.
The dressed states' analysis pinpointed several cases where constructive and destructive interference between HF excitation pathways in a two-photon excitation scheme limits the available two-photon transitions. 
This results in a reduction of the conventional two-photon selection rule for the total angular momentum $F$, from $\Delta F= 0,\pm 1$ to $\Delta F\equiv 0$.
Our discovery presents practical methods for selectively controlling the populations of unresolvable HF $F$-components of $ns_{1/2}$ Rydberg states in alkali metal atoms.
Using numerical simulations with sodium and rubidium atoms, we demonstrate that by blocking the effects of HF interaction with a specially tuned auxiliary control laser field, the deviations from the ideal selectivity of the HF components population can be lower than $0.01\%$ for Na and $0.001\%$ for Rb atoms.
\end{abstract}

% PACS, the Physics and Astronomy Classification Scheme.
 %Use showpacs class option if PACS display desired
% 33.70.Jg Line and band widths, shapes, and shifts
% 32.70.Jz Line shapes, widths, and shifts
% 31.15.-p Calculations and mathematical techniques in atomic and molecular physics

%\keywords{transit time broadening, excitation spectra, two-photon excitation, lineshapes, linewidths}%Use showkeys class option if keyword display desired

\maketitle

%-------------------------{introduction} begin-----------------------------

\section{\label{introduction}Introduction}

Development of Quantum Computing (QC) \cite{Ladd2010}, one of the rapidly emerging modern technologies, faces several critical challenges in producing a reliable quantum computing device.
Coupling to the environment and the generally high sensitivity of quantum systems to external influences lead to decoherence, measurement noise and computation errors.
In fault-tolerant approaches to QC, the majority of physical qubits are used by error correction \cite{PhysRevA.86.032324}, where they have to be initialized and read out repeatedly during the computation. 
Development of high-fidelity qubit manipulation techniques can alleviate the resource-heavy error correction by reducing the error rates.
Preparation of quantum objects into a desired state seems to be one of the today’s biggest challenges \cite{Saffman_2016}.
Indeed, we can only trust the results of computations when qubits are consistently initialized exactly in the required state.
The preferred preparation process depends on the physical implementation of the qubit. 
In the case of neutral atoms, coherently controlled electronic excitation into a Rydberg state is often preferred as it allows to manipulate the excited state conveniently \cite{Saffman2010,Ryabtsev2016}.

%---------------Fig1(5)-beg-------
\begin{figure}
    % \centering
    \includegraphics[width=\linewidth]{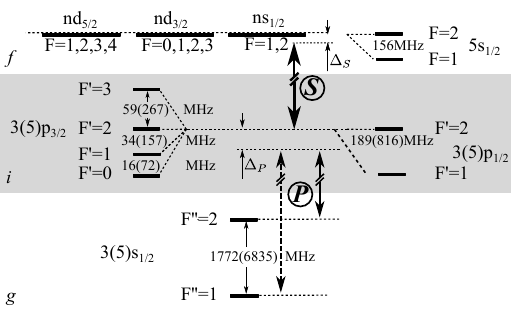}
    \caption{
    Excitation schemes in $^{23}$Na (and $^{87}$Rb, the numbers in parentheses).
    The weak probe laser $P$ excites the transition between the ground state $g$ and the intermediate state $i$, whereas the strong field $S$ couples the intermediate state $i$ to the final state $f$, forming the adiabatic (laser-dressed) states.
    The detunings $\Delta_{S,P}$ of the laser fields with the frequencies $\omega_{S},\omega_{P}$ are defined relative to the resonance frequencies of the respective HF transitions between components with $F'=2$ and $F=2$.
    For instance, $\Delta_{S}=\omega_{S}-\omega_{5s(F=2), 3p(F'=2)}$, while $\Delta_{P}=\omega_{P}-\omega_{3p(F'=2), 3s(F'')}$, so that $\Delta_{S,P}$, as shown in the schematic, have negative signs.
    }
    \label{fig5}
\end{figure}
%---------- Fig1(5)--end------

Although a variety of coherent techniques such as the stimulated Raman adiabatic passage technique (STIRAP) \cite{Vitanov2017,Shore2017} have enabled nearly $100\%$ efficient excitation of the selected target quantum state \cite{PhysRevLett.95.043001}, for alkali-metal atoms the perfect level addressing problem remains unsolved due to existence of Hyperfine (HF) splitting of the atoms’ energy levels.
HF structure of highly excited levels with large principal quantum numbers $n$ can not be resolved due to the small energy separation ($\sim n^{-3}$) between sublevels.
Thus no separate HF sublevel can be populated by any tuning of the exciting laser.
Our research presented here (see Sections \ref{Architecture} and \ref{ManipulationsC}) demonstrates that selective excitation can still be achieved for a few two-photon excitation schemes of type $3(5)s_{1/2}- 3(5)p_{3/2,1/2} - ns_{1/2}$ in $^{23}$Na and $^{87}$Rb atoms (see Fig.~\ref{fig5}).
Throughout this paper, the principal quantum numbers and the HF energy level splitting values placed in parentheses refer to rubidium atoms.
In these excitation schemes, the total angular momentum ${F}''$ of a well-resolved ground $3(5)s_{1/2}$ state has to remain the same for the final $ns_{1/2}$ state.
As long as the P-laser couples all HF components $F'$ of a $3(5)p_{3/2,1/2}$ state in the first excitation step ($\Delta F_{P S}= 0,\pm 1$ according to the one-photon selection rule), while in the second step the S-laser couples both HF components $F=1,2$ of $ns_{1/2}$ state ($\Delta F_{S P}= 0,\pm 1$), one of which turns out to be unpopulated, we can speak of a specific two-photon selection rule $\Delta F_{ SPS}\equiv 0$ observed instead of the expected $\Delta F_{SPS}= 0, \pm 1$.

We study the emergence of this ``modified'' selection rule for certain two-photon transitions in alkali metal atoms, focusing on sodium and rubidium due to their widespread use in applied fields of physics.
As will be shown later, the set of suitable excitation schemes may be extended to $rs_{1/2} - kp_{3/2,1/2} - ns_{1/2}$, where the principal quantum numbers $k\geqslant r$; $n > k$.
For our purpose, we will employ an Autler-Townes (AT) spectroscopic experiment as it allows observing populations of the intermediate ($i$) and the highest (final, $f$) states of a 2-step excitation scheme \cite{PhysRev.100.703}.
In a typical AT arrangement (see Fig.~\ref{fig5} for the corresponding energy diagram), a strong (S) laser couples intermediate and final levels, producing adiabatic (``laser-dressed'') states, while a weak probe (P) laser couples ground ($g$) and intermediate levels, providing a modest population of the adiabatic states.

%---------Fig2(figSh)---beg------
%------------------------------
\begin{figure}
    % \centering
    \includegraphics[width=\linewidth]{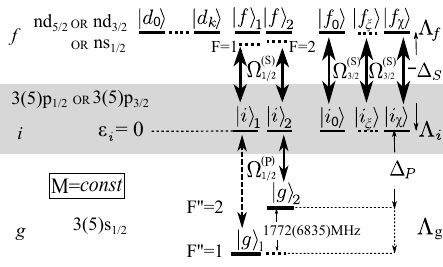}
    \caption{
    The same level scheme as in \figref{fig5} but under RWA in the semi-hyperfine basis of the Morris-Shore type with broken HF coupling in the intermediate i-subspace.
    The sets of Bright $\ket{i,f}_{1,2}$, Chameleon $\ket{i,f_{0 \dots \chi}}$ and Dark $\ket{d_{0\dots k}}$ states in the $i$- or $f$-spaces $\Lambda_f$ or $\Lambda_i$ depend on the magnetic quantum number $M$ and on the $f$-, $i$-states' quantum numbers $F$, ${F}'$ respectively.
    The index $\eta=1,2$ that is used with states $\ket{g,i,f}_{\eta}$ denotes the first component of the double-index $\eta \!=\!F''M$ (see its definition in Sec.~\ref{MSbasis}) while the second component $M$ is shown on the figure itself.
    %In the current example it indicates only the quantum number $F''$.
    %For example, for Zeeman sublevels with $M=1$ there is one Chameleon ($\chi =0$) and one Dark ($k=1$) state, while $\ket{g,i,f}_{1} \equiv \ket{g,i,f}_{1,1}$, $\ket{g,i,f}_{2} \equiv \ket{g,i,f}_{2,1}$.
    The RWA energies $\varepsilon_i$ of $i$-states are chosen as zero, while those $\varepsilon_g = \Delta _P$, $\varepsilon _f (F=2)= -\Delta _S$ of $g$- and $f$-states are determined by laser detunings $\Delta _{P,S}$.
    }
    \label{figSh}
\end{figure}
%-----Fig2(figSh)---end-----
%------------------------------

The simultaneous interaction of multiple HF and Zeeman components leads to a complicated and difficult to analyze excitation spectrum, due to the presence of a number of different Rabi frequency values. 
Authors of work \cite{PhysRevA.27.906} developed the so-called Morris-Shore (MS) transformation for finding a special set of basis wave vectors (MS basis), which reduces a coupled two-level system with degenerate sublevels (Zeeman sublevels for instance) to a set of coupled ``bright'' pairs (BS) and single decoupled ``dark'' states (DS) (see also in \cite{Bevilacqua2022}).
Noticeable HF splitting (especially in Rb) introduces a fundamental limitation for practical use of the MS method.
However, if the S-laser coupling is much stronger than the HF interaction between sublevels (see $i \rightarrow f$ transition in Fig.~\ref{figSh}), then adiabatic states are formed by pairs of coupled BS ($\ket{i,f}$ states in Fig.~\ref{figSh}) and by a set of single noninteracting DS ($\ket{d}$ states in Fig.~\ref{figSh}).
As it was shown in works \cite{Kirova2017,Cinins2021}, some BS (termed ``Chameleon'' states) acquire features of dark states provided that their ($\ket{f_{0}}, \dots ,\ket{f_{\chi}}$ states in Fig.~\ref{figSh}) are not directly coupled to the ground state.

The fluorescence radiation of adiabatic states bears information about the states’ populations, and measuring it while scanning the P-laser frequency, yields AT-spectrum that reveals a set of levels: the fluorescence peak positions and heights correspond to energies and populations of the adiabatic states (see the corresponding spectra in Sec.~\ref{AT}).
Due to the large HF splitting in the ground state, the adiabatic states are probed from either the $F''=1$ or $F''=2$ HF level of the $g$-state.
The key result of our work is that upon probing, one is in fact addressing independent (orthogonal) sets of adiabatic states representing separate three-level ladder-like sequences: one set is selectively excited from $F''=1$, and a different set from $F''=2$ (Fig.~\ref{figSh}).
If the selection rule $\Delta F\equiv 0$ is satisfied for the final $ns_{1/2}$ state, then only one $F$-sublevel, namely $F=F''$, is populated and the AT spectrum will contain only two peaks that correspond to a single bright states pair.
And indeed, the spectrum associated with the $3s_{1/2} - 3p_{3/2} - 5s_{1/2}$ sequence in Na is found to be just as expected (see Fig.~\ref{Selective_3/2} in Sec.~\ref{AT}).

In our previous works \cite{Kirova2017,Cinins2021}, devoted to numerical experiments with sodium atoms, we performed calculations of AT spectra for two of the excitation schemes, namely $3s_{1/2}-3p_{3/2} -4d_{5/2,3/2}$.
It was found there empirically, that, after a special, MS-like transformation in HF sublevels of the intermediate $3p_{3/2}$ state, the excitation scheme is reduced to a set of independent simple $\text{three-,}$ two- and single-level blocks (as shown in \figref{figSh}), provided that HF splittings of intermediate i-states may be ignored.
In this work, we demonstrate analytically two fundamental facts which hold for any alkali metal atom.
(i) The simplification revealed in \cite{Kirova2017,Cinins2021} is a characteristic feature (Sec.~\ref{Architecture}) of all linkage diagrams in the case of linear polarization of the P- and S-laser fields.
(ii) The two-photon selection rules $\Delta F =0$ are satisfied regardless of the P- and S-laser intensities, provided that the last ladder step is $ns_{1/2}$ state and the proper elimination of HF interaction at the step $i$ is performed using special manipulation instruments, such as an additional control laser and properly adjusted laser detunings (Subsec.~\ref{ManipulationsC} and Appendix~\ref{App3}).
The physical reasons for our findings are associated with the specific properties of linearly polarized laser fields, namely, with the fact that the corresponding operators of their interaction with atomic states turn out to be semi-unitary in a sense, partially retaining the orthogonality property when acting on the atomic wave functions.
The exact formulation and algebraic proof of this fact are given in subsection ~\ref{Matrix} and in Appendix~\ref{App0}.
Noteworthy, linearly polarized lasers are widely used to create optical dipole traps due to uniform ac Stark (light) shifts that do not depend on the magnetic quantum numbers $M$ \cite{Grimm2000,Hofmann2013}.

The rest of paper is organized as follows.
Subsection~\ref{MSbasis} is devoted to the construction of a special MS basis of wave functions at each step of ladder excitation, demonstrating that the HF operator turns out to be diagonal (or negligible for $nd$ states) in the basis of the ground and final steps.
To take into account HF interaction at the intermediate i-step, the matrix form of the HF operator is provided (Subsec.~\ref{HFI}), which makes it possible to estimate both HF-splitting and HF-mixing for elements of the MS i-basis.
The latter is a key for finding the necessary parameters for the auxiliary laser, which is used (Subsec.~\ref{ManipulationsC}) to partially block HF mixing within MS $i$-basis and, thus, to control the two-photon selection rules.
In Sec.~\ref{AT} after a brief discussion of the basic terms and a formalism through which Bright, Chameleon and Dark states manifest themselves in fluorescence, we will illustrate our theoretical findings with numerically calculated AT spectra for a few excitation schemes at hand.
The mechanism underlying the AT spectra reduction and its connection with selection rules is described in the next Sec.~\ref{selectivity}.
Following the method used in the phenomenon of electromagnetically induced transparency \cite{Fleischhauer2005}, Subsec.~\ref{ManipulationsC} demonstrates how the correct tuning of an auxiliary control laser can bypass the limitations imposed by the HF interaction effects on the degree of selectivity upon excitation of the HF components of Rydberg $ ns_ {1/2} $ states.
The paper ends with a conclusion and acknowledgments.
Appendix~\ref{App0} gives mathematical formulations of the field operators semi-unitarity, while Appendix~\ref{App1} gives survey of the numerical technique and atomic data used in our AT spectra calculations.
Finally, Appendix~\ref{App3} is devoted to assessments of the selectivity factors depending on parameters of the control laser used for blocking  the HF effects.

%For those readers who are concerned with the applied aspects of the dressed states' formalism, we recommend skipping Section~\ref{Architecture} burdened with mathematical details and turning directly to Sections~\ref{AT}, \ref{selectivity}.

Atomic units are used throughout this paper unless stated otherwise.

\section{HF blocks architecture of two-step excitation schemes}
\label{Architecture}

Dynamics of atomic systems under the influence of periodic or polychromatic control fields can be accurately described using various modifications of the Floquet technique \cite{Chu2004,efimov2014,Eckardt2015}.
A useful tool for theoretical analysis in the case of nearly–resonant monochromatic laser fields $\textbf{E}\cos(\omega _Lt)$ is the rotating wave approximation (RWA) \cite{Shore2009}.
In this framework, the problems of light-matter interactions are reduced to solving the quasistationary optical Bloch equations for evolution of the atomic density matrix, with characteristic timescales being determined by the duration of laser pulses, i.e. the temporal behavior of the electrical field amplitudes $\textbf{E}(t)$.
A key component of the theoretical analysis is the concept of adiabatic (``laser-dressed'') atomic state basis, which often allows to develop quantitative description of the phenomena at hand for relatively simple systems and to predict possible evolution scenarios in the case of multidimensional configurations \cite{Shore2017,Rangelov2006}.
The latter have been considered as a promising physical objects for quantum processing on the so-called qudits (multidimensional extension of qubits), permitting to design effective algorithms for the implementation of universal gates via adiabatic passage processes \cite{Rousseaux2013,Cinins2022}.
In this section we develop an approach that enables identifying dressed states in multidimensional two-step ladder excitation schemes, in the specific case of alkali metal atoms interacting with linearly polarized laser light.

\subsection{Notation and assumptions}
\label{Notations}

Under the framework of RWA, the adiabatic states $\ket{\alpha}$ are obtained as eigenfunctions $\operator{H}_{RW}\ket{\alpha} =\alpha \ket{\alpha}$ of the quasistationary RWA Hamiltonian
\begin{align}
    \label{eq:Hamiltonian}
    \operator{H}_{RW}\!=\!\operator{H}^{(a)}\!+\!
    \operator{H}^{(hf)}\!+\!\operator{H}^{(ph)}; \quad
    \operator{H}^{(ph)}\!=\!
    \operator{V}^{(S)}\!+\!\operator{V}^{(P)}
\end{align}
\noindent which describes interaction of atomic systems with external electromagnetic fields.
The Hamiltonian $\operator{H}^{(a)}$ of intra-atomic interactions up to the fine structure interaction, together with the HF operator $\operator{H}^{(hf)}$ determine the energy structure of the diabatic (bare) states of an isolated atom.
The field operators
\begin{align}
    \label{Field}
    \operator{V}^{(S,P)} \!=\!-\frac{1}{2}\operator{\textbf{d}}
    \textbf{E}^{(S,P)}
\end{align}
\noindent describe the coupling of the atomic dipole moment $\operator{\textbf{d}}$ with the slowly varying amplitudes $\textbf{E}^{(S,P)}$ of linearly polarized S- and P-laser fields.
The direction of the quantization axis is chosen along the polarization unit vector $\textbf{e}_z$ common to both laser fields, i.e. $\textbf{E}^{(S,P)}=E^{(S,P)}\textbf{e}_z$.
Importantly, the magnetic quantum numbers $M$ are dynamic invariants due to the azimuthal symmetry of the ``atom + laser fields'' system \cite{Landau1981}.
This feature allows us to treat each subset of mutually optically coupled HF levels with the same $M$ as an independent multilevel system $\Lambda_M$.

Matrix representation of the field operators in a specifically chosen quantum states basis reveals important algebraic properties, that are instrumental for studying the structure of the adiabatic states.
For all two-step excitation schemes in alkali metal atoms depicted in Fig.~\ref{fig5} and their analogues, the field operators act on the wave vectors space $\Lambda=\Lambda_{g} \oplus \Lambda_{i} \oplus\Lambda_{f}$, which is a direct sum of subspaces $\Lambda_{\gamma}$, corresponding to the three ${\gamma =g,i,f}$ ladder steps and consisting of Zeeman and HF sublevels $\ket{\gamma,F,M}$.
We sometimes omit the symbols of the quantum numbers in wave vector notation, indicating that the vectors belong to a specific ladder step by the parameter $\gamma = g, i, f$ equivalent to $\gamma = ks_{1/2}, kp_{1/2,3/2}, \{ns_{1/2}, nd_{3/2,5/2}\}$, respectively.
Although the operators $\operator{V}^{(\Im )}$ ($\Im = S,P$) are self-adjoint, it is beneficial to use the following notation
\begin{equation}
    \begin{split}
        \operator{V}^{(S)}\Lambda_{i}\!\to\!\Lambda_{f}; \quad \operator{V}^{(S)\dagger}\Lambda_{f}\!\to\!\Lambda_{i}; \\ \operator{V}^{(P)}\Lambda_{g}\!\to\!\Lambda_{i}; \quad \operator{V}^{(P)\dagger}\Lambda_{i}\!\to\!\Lambda_{g},
        \label{Flip}
    \end{split}
\end{equation}
\noindent implying that the laser-atom interactions should be interpreted as mapping operations between excitation manifolds.
The operator $\operator{V}^{(S)}$ ($\operator{V}^{(P)}$) maps the intermediate subspace $\Lambda_ {i}$ (the ground subspace $\Lambda_ {g}$) onto the final subspace $ \Lambda_ {f}$ (the intermediate subspace $\Lambda_ {i}$) and vice versa for the conjugated operator $\operator {V}^{(S)\dagger }$ ($\operator {V}^{(P)\dagger }$).
Remarkably, the mappings~(\ref{Flip}) partially preserve orthogonality of wave vectors, i.e. they can be called semi-unitary.
We expand on this property in the following subsections.
A more rigorous formulation of a semi-unitary operator provided in Appendix~\ref{App0} will allow us to find in Sec.~\ref{MSbasis} an algorithm for constructing a special Morris-Shore basis shown in Fig.~\ref{figSh}.

\subsection{Dipole matrix elements}
\label{Matrix} 

The dipole matrix elements of the field operators~(\ref{Field}) are associated with their Rabi frequencies \cite{Sobelman1992}
\begin{equation}
    \begin{split}
        \bra{\gamma ,F,M} 2\operator {V}^{(S) }
        \ket{\gamma ',F',M'}\equiv ^{(S)}\!\!\Omega_{\gamma 'F'}^{\gamma
        F}(M)\delta _{MM'}; \\
        \bra{\gamma '\!,F'\!,M'} 2\operator {V}^{(P) }
        \ket{\gamma ''\!,F''\!,M''}\!\equiv ^{(P)}\!\!\Omega_{\gamma 'F'}^{\gamma
        ''F''}(M)\delta _{M'M''}
        \label{Rabi_HF}
    \end{split}
\end{equation}
\noindent
which are presented in Appendix~\ref{App1} in the HF-basis $\ket{\gamma ,F,M}$ (see Eqs.~(\ref{rabi1}), (\ref{rabi2})), helpful for numerical modeling of AT spectra.
Algebraic features of the operators $\operator{V}^{(S,P)}$ for linearly polarized laser fields, along with their spectral properties, are more naturally displayed in the fine strucutre (FS) basis.
At each ladder step $\gamma$ it consists of the product $|\gamma ,lJm_{J}m_{I}\rangle\!=\! |\gamma ,lJm_{J}\rangle |m_{I}\rangle$ of basis vectors of electron $|\gamma ,lJm_{J}\rangle$ and nuclear spin $|m_{I}\rangle$.

\begin{figure}
    \centering
    \includegraphics[width=\linewidth]{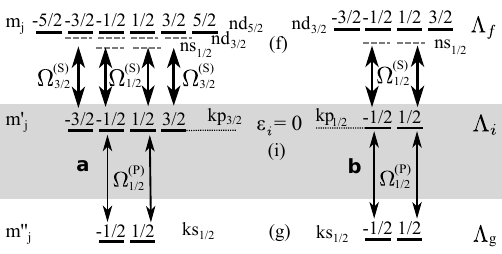}
    \caption{
     Two-photon ladder excitation schemes in the fine structure basis of alkali metal atoms, without taking into account the HF interaction.
    Under RWA, the energies $\varepsilon _{\gamma}$ of the $\gamma$-states ($\gamma =g,i,f$) are determined by the lasers detunings $\Delta _{P,S}$, in the same manner as in Fig.~\ref{figSh}.
    The principal quantum number $k$ of the ground states is $3$ for Na and $5$ for Rb atoms, respectively.
    }
    \label{figbreak}
\end{figure}

The main advantage of the FS $\gamma$-basis is that it presents the natural Morris-Shore bases for the field operators $\operator {V}^{(\Im)}$ provided that we entirely ignore the HF interaction (see Fig.~\ref{figbreak}).
This statement follows from the corresponding Rabi frequencies of the optical transitions $\{\gamma , lJm_{J}m_{I}\to \tilde{\gamma }, \tilde{l}\tilde{J}\tilde{m}_{J}\tilde{m} _{I}\}$ \cite{Sobelman1992}
\begin{multline}
    ^{(\Im)}\Omega_{\tilde{\gamma }, \tilde{l}\tilde{J}}^{(\gamma ,lJ)}(m_{J}) \!=\!(-1)^{\Phi-m_{J}} \Omega_{\Im }
    \left\lbrace \begin{matrix} l & J & s \\ \tilde{J} & \tilde{l} & 1 \end{matrix} \right\rbrace
    \times
    \\
    \sqrt{(2\tilde{J}+1)(2J+1)}
    \left( \begin{matrix} J & 1 & \tilde{J} \\ -m_{J} & 0 & m_{J} \end{matrix} \right)\delta _{m_{J}\tilde{m}_{J}}\delta _{m_{I}\tilde{m}_{I}},
    \label{rabiFS}
\end{multline}
\noindent
where linear laser polarizations imply that magnetic quantum numbers $\tilde{m}_{J}=m_{J}$. 
Optical transitions do not affect the nuclear spin variables, therefore also $\tilde{m}_{I}=m_{I}$.
The phase $\Phi\!=\!\tilde{l}\!+\!s\!+\!1\!+\!\tilde{J}\!+\!J$ incorporates the electron spin $s\!=\!1/2$ and orbital $l$ quantum numbers, the index {$\Im $} stands for the $P$- or $S$-laser excitation, $m_{J}\!=\!-J,-J+1,\dots,J$, and $m_{I}\!=\!-I,-I+1,\dots,I$ (pay attention that in the HF basis $M=m_J+m_I$).
The non-essential for the present discussion multipliers $\Omega_{\Im } $ are the reduced Rabi frequencies, determined by Eq.~(\ref{rabi1}) in Appendix~\ref{App1}.

%--------------------------

%----Fig3(figbreak)-beg---------
%------------------------------
%---Fig3(figbreak)-end---
%------------------------------

The relation (\ref{rabiFS}) indicates that both laser fields couple only the levels with the same quantum numbers $m_{J},m_{I}$, and the corresponding linkage diagram shown in Fig.~\ref{figbreak} is reduced to a set of separate, non-interacting (independent) three-level ladders
\begin{align}
    \label{LaddersFS}
    \ket{ks_{1/2}m_J}\! \to \! \ket{kp_{3/2,1/2}m_J} \! \to \! \ket{nd_{5/2,3/2}, ns_{1/2}m_J},
\end{align}
\noindent similar to those depicted in \figref{figSh}.
Importantly, Eq.~(\ref{rabiFS}) determines the effective Morris-Shore Rabi frequencies (MSF) $\Omega_{\chi}^{{eff}}$ \cite{PhysRevA.27.906} of the operators $\operator{V}^{(\Im)}$:
\begin{align}
    \label{RabiEf}
    \Omega_{\chi}^{{eff}}=\Omega_ {m_{J}}^{(\Im)} \!\equiv\! \frac{1}{2} \left |^{(\Im)}\Omega_{\tilde{\gamma } \tilde{L}\tilde{J}}^{(\gamma ,lJ)}(m_{J}) \right |
\end{align}

If we change the sign of $m_j$, then, due to the equality \cite{Sobelman1992}
\begin{equation}
    \left( \begin{matrix} J & 1 & \tilde{J} \\ -m_{j} & 0 & m_{j} \end{matrix} \right) = (-1)^{J+\tilde{J}+1} \left( \begin{matrix} J & 1 & \tilde{J} \\ m_{j} & 0 & -m_{j} \end{matrix} \right)
    \label{RabiOpo}
\end{equation}
\noindent 
and Eqs.~(\ref{rabiFS}), (\ref{RabiEf}), it follows that $\Omega_{-m_{J}}^{(\Im)}\!=\!\Omega_ {m_{J}}^{(\Im)}$.
In other words, the coupling between subspaces $\Lambda_g,\Lambda_i$ has a singular effective MSF value $\Omega_ {1/2}^{(P)}$, while for the $i-f$ coupling there are two possible MSF values
\begin{align}
    \label{RabiEfS}
    \Omega_{1/2,3/2}^{(S)}= \frac{1}{2} \left | ^{(S)} \Omega_{\tilde{\gamma }, \tilde{L}\tilde{J}}^{(\gamma ,lJ)}(m_{J}=1/2,3/2) \right |,
\end{align}
\noindent
depending on whether $|m_{J}|=1/2$ or $3/2$.

\subsection{Building the Morris-Shore basis vectors }
\label{MSbasis} 

Although Eq.~(\ref{RabiEf}) is obtained for a specific fine structure basis, it can be expressed in an alternative invariant operator form related to important semi-unitarity features of the field operators: $\operator{V}^{(P,S)}$ provide the mappings~(\ref{Flip}) of subspaces $\Lambda _{\gamma}$ while maintaining fully (in the case $\gamma \! =\! i\! =\! kp_{1/2}$) or partially ($\gamma \! =\! i\! =\! kp_{3/2}$) orthogonality of wave vectors.
The corresponding mathematical discussion is given in Appendix~\ref{App0}, with a statement of what semi-unitarity means in the paragraph immediately after Eq.~(\ref{DotProduct_i}).

That semi-unitarity enables one to reduce the complex excitation diagrams for alkali metal atoms (of the type shown in Fig.~\ref{fig5}) to combinations of simple ladder excitation schemes due to applying two sequential semi-unitary mappings of ground HF sublevels:
\begin{equation}
    \begin{split}
        \qquad \hat{V}^{(P)}|g,F''M \rangle & \to \Omega _{1/2}^{(P)} \cdot|i\rangle_{F''M};
        \\
         \hat{V}^{(S)}|i\rangle_{F''M} & \to \Omega _{1/2}^{(S)}\cdot|f\rangle_{F''M}
    \end{split}
    \label{mapping}
\end{equation}
\noindent Here $|g,F''M \rangle \!=\! |g\rangle_{F''M}$ is the initial diabatic vector from the HF g-basis (see in Fig.~\ref{figSh}), while the unit vectors $|i\rangle_{F''M}$ and $|f\rangle_{F''M}$ are its images in the $\Lambda _{i}$ and $\Lambda _{f}$ subspaces.
Since two vectors $|g\rangle_{\eta, \widetilde{\eta}}$ with different double-indexes $\eta \!=\!F''M $ and $\widetilde{\eta} \!=\!\widetilde{F}''\widetilde{M} $ are orthogonal, their images $|i\rangle_{\eta, \widetilde{\eta}}$ and $|f\rangle_{\eta, \widetilde{\eta}}$ at both i- and f-excitation steps should be mutually orthogonal as well and have unit magnitudes in accordance with Eqs.~(\ref{DotProduct}), (\ref{Orthogonalg_i}), (\ref{DotProduct_i}).
Therefore, each three-step ladder (\ref{mapping}) represents a unique and independent (orthogonal to others) excitation path, predefined by the choice of the index $\eta $, with the effective MS Rabi frequencies $\Omega ^{(P,S)}_ {1/2}$ related to the P- and S-lasers.

The above ladder basis levels $|\gamma\rangle _{\eta}$ ($\gamma = i, f$), being directly coupled to the ground states (see Fig.~\ref{figSh}), constitute the manifolds $\Lambda _{\gamma }^{BS}$ of bright states in the subspaces $\Lambda _{\gamma }$ (see Table~\ref{tab00}).
Since $\Lambda _{i}^{BS}=\hat{V}^{(P)}\Lambda _{g }$, then $\Lambda _{i}^{BS}=\Lambda _{i,1/2 }$ (see Eq.~(\ref{Image})), where the manifolds $\Lambda _{i,\lambda }$ are defined after Eq.~(\ref{Prjections}) as consisting of all eigenvectors corresponding to the
eigenvalue $|\Omega _{\lambda}^{(S)}|^2 $ of the operator $\hat{V}^{(S)\dagger}\hat{V}^{(S)}$.
It's obvious that $\Lambda _{f }^{BS}=\hat{V}^{(S)} \Lambda _{i }^{BS}$.

%-----------------------------Tab1-beg-----------------------------
\begin{table}%[H]
    \caption{The structure of subspaces $\Lambda _{\gamma}$ of wave vectors, related to three g-, -i, f-steps of the ladder excitation scheme, depending on the i-f configurations of the ladder.}
    \label{tab00}
    \begin{tabular}{|c|c|c|c|}
        \hline
        Configuration $i-f$ & $\Lambda _g$ & $\Lambda _i$ & $\Lambda _f$
        \\
        \hline
        $1)\; kp_{1/2}-ns_{1/2}$ & $\Lambda _g^{BS}$ & $\Lambda _i^{BS}$ & $\Lambda _f^{BS}$ \\
        %\hline
        $2)\; kp_{1/2}-nd_{3/2}$ & $\Lambda _g^{BS}$ & $\Lambda _i^{BS}$ & $\Lambda _g^{BS} \oplus \Lambda _f^{DS}$ \\
        %\hline
        $3)\; kp_{3/2}-ns_{1/2}$ & $\Lambda _g^{BS}$ & $\Lambda _i^{BS} \oplus \Lambda _i^{DS}$ & $\Lambda _f^{BS}$ \\
        %\hline
        $4)\; kp_{3/2}-nd_{3/2}$ & $\Lambda _g^{BS}$ & $\Lambda _i^{BS} \oplus \Lambda _i^{CS}$ & $\Lambda _f^{BS} \oplus \Lambda _f^{CS}$
        \\
        %\hline
        $5)\; kp_{3/2}-nd_{5/2}$ & $\Lambda _g^{BS}$ & $\Lambda _i^{BS} \oplus \Lambda _i^{CS}$ & $\Lambda _f^{BS} \oplus \Lambda _f^{CS} \oplus \Lambda _f^{DS}$ \\
        \hline
    \end{tabular}
\end{table}
%-----------------------------TabA1-end-----------------------------

In the case of \figref{figbreak}(a) ($i\!=\!3p_{3/2}$), the intermediate $\Lambda _i $ subspace includes another manifold $\Lambda _{i,3/2}$, orthogonal to $\Lambda _{i,1/2}$.
We can choose in $\Lambda _{i,3/2}$ a complete set of basis vectors $|i_{\xi M'}\rangle$ ($\xi = 0,1,...,\chi$), where the the number $\chi$ of possible values of the integer index $\xi $ depends on the magnetic quantum number $M'$. Tables~\ref{tab21_tab22},\ref{tab23_tab24} shows the possibles sets of $|i_{\xi M'}\rangle$ with different $M'$ using Na and Rb as an example.
Note that the number $M'$ is excluded from the notation of wave vectors if $M'$ is explicitly indicated, as it done in captions to Table~\ref{tab21_tab22} and to Fig.~\ref{figSh}.

As mentioned above, the set of vectors $|f\rangle _{F''M}$ (\ref{mapping}) constitute the manifold $\Lambda _f^{BS}$ of bright states in the final ladder subspace $\Lambda _f$.
If $f\!=\!ns_{1/2}$ then $\Lambda _f\!=\!\Lambda _f^{BS}$.
Otherwise ($ f \!= \!nd $) we should supplement the bright f-vectors by maps of the basis vectors from the manifold ${\Lambda _{i,3/2}}$:
\begin{equation}
    \hat{V}^{(S)}|i_{\xi M }\rangle \to \Omega _{3/2}^{(S)}\cdot|f_{\xi M}\rangle
    \label{Fns}
\end{equation}
\noindent Due to the semi-unitarity (\ref{DotProduct_i}) of the operator $ \hat{V}^{(S)} $, the vectors $|f_{\xi M}\rangle $ along with bright vectors $|f\rangle _{F''M}$ have two main properties of their preimages $|i_{\xi M}\rangle $, $|i\rangle _{F''M}$: (1) they have unit length and (2) they are all mutually orthogonal.

Note, that, formally, the basis vectors entering Eq.~(\ref{Fns}) are involved in the laser-atom interaction with the effective MS Rabi frequency $\Omega _{3/2}^{(S)}$.
For this reason they may be called ``Bright''.
On the other hand, it is impossible to excite $|i_{\xi M}\rangle $ or $|f_{\xi M}\rangle $ directly from the ground states (see \figref{figSh}).
Therefore, they should belong to the class of ``Dark'' \cite{PhysRevA.27.906}. In other words, the basis vectors $|i_{\xi M}\rangle $, $|f_{\xi M}\rangle $ share the features of both ``Bright'' and ``Dark'' states, which gave reason to assign them the name ``Chameleons'' \cite{Kirova2017} and associate the manifolds $\Lambda _{i,f,3/2}$ with subspaces $\Lambda _{i,f}^{CS}$ of Chameleon states.
As a consequence, $ \Lambda _{i,f} \!= \!\Lambda _{i,f}^{BS}\oplus \Lambda _{i,f}^{CS}$ (see Figs.~\ref{figSh}, \ref{figbreak} ) as it indicated in row 4) of Table~\ref{tab00}.

Note, however, that for the ladder configuration corresponding to raw 3) of the Table, the value $\Omega _{3/2}^{(S)}=0$ (see Eq.~(\ref{RabiOpo})) and, according to Eq.~(\ref{Fns}), $\Lambda _{f}\!=\!\Lambda _f^{BS}$.
In other words, the Chameleon states $|i_{\xi M}\rangle$ are not associated here with any laser interaction and should be treated as ``Dark''.

The configurations in rows 2) and 5) of Table~\ref{tab00} require additional consideration.
In the case of row 5), the direct sum $\Lambda _{nd_{5/2}}^{BS}\oplus \Lambda _{nd_{5/2}}^{CS}$ is part of the the subspace $\Lambda _{nd_{5/2}}$ (see \figref{figbreak}).
Therefore, it has an orthogonal complement $\Lambda _{nd_{5/2}}^{DS}$, such that
\begin{equation}
    \Lambda _{nd_{5/2}}=\Lambda _{nd_{5/2}}^{BS}\oplus \Lambda _{nd_{5/2}}^{CS} \oplus \Lambda _{nd_{5/2}}^{DS}
    \label{DS}
\end{equation}
\noindent The arbitrarily chosen basis vectors $|d_{M \xi}\rangle$ ($\xi \!=\!0,...,\kappa$) in the manifold $\Lambda _{nd_{5/2}}^{DS}$ are in no way coupled with either the ground or intermediate states (see \figref{figSh}), i.e. are ``Dark''.

In the case of configuration 2) in Table~\ref{tab00}, when subspace $\Lambda _{f}$ consists of the direct sum of the subspaces $\Lambda _{f}^{BC}$ and the orthogonal complement $\Lambda _{f}^{DC}$ to it, the basis vectors $|d_{M \xi}\rangle$ from $\Lambda _{f}^{DC}$ also belong to the ``Dark'' class.

\subsection{\label{HFI} Accounting for Hyperfine interaction}

%Tab1-%beg--------Na-1-----------------
\begin{table}%[H]
    \caption{
    Matrix elements of the HF operator in the MS basis of $\Lambda_i$ subspaces (a)~$3p_{3/2}$ and (b)~$3p_{1/2}$ for manifolds with different $M$, in units of MHz for the case of Na atoms. The diagonal elements give the shifts $\varepsilon_{i}$ of the initial RWA energies $\varepsilon_{i} \equiv 0$ resulted due to the HF interaction, while the off-diagonal elements are Rabi frequencies $\Omega _{HF}$ of HF mixing between MS states.
    }
    \label{tab21_tab22}
    \begin{minipage}[t]{0.60\linewidth}
    % \centering
    % \caption*{(a)}
    \begin{center}
    (a)~$3p_{3/2}$ manifold\\
    % \label{tab21}
    $M=0$\\
    \begin{tabular}{|c|cccc|}
        \hline
        States & $\ket{i}_1$ & $\ket{i}_2$ & $\ket{i_{0}}$ & $\ket{i_{1}}$ \\
        \hline
        $\ket{i}_1$ & -25.08 & 0.000 & 25.08 & 0.000 \\
        %\hline
        $\ket{i}_2$ & 0.000 & 49.06 & 0.000 & 27.80 \\
        %\hline
        $\ket{i_{0}}$ & 25.08 & 0.000 &-25.08 & 0.000 \\
        $\ket{i_{1}}$ & 0.000 & 27.80 & 0.000 & -25.08 \\
        \hline
    \end{tabular} \vspace{5pt}\\
    $M= \pm 1$ \\
    \begin{tabular}{|c|ccc|}
        \hline
        States & $\ket{i}_1$ & $\ket{i}_2$ & $\ket{i_{0}}$ \\
        \hline
        $\ket{i}_1$ & -21.47 & -7.44 & 14.87 \\
        %\hline
        $\ket{i}_2$ & -7.44 & 44.09 & 28.48 \\
        %\hline
        $\ket{i_{0}}$ & 14.87 & 28.48 & 1.36 \\
        \hline
    \end{tabular} \vspace{5pt}\\
    $M= \pm 2$\\
    \begin{tabular}{|c|cc|}
        \hline
        States & $\ket{i}_2$ & $\ket{i_{0}}$ \\
        \hline
        $\ket{i}_2$ & 29.16 & 29.16 \\
        %\hline
        $\ket{i_{0}}$ & 29.16 & 29.16 \\
        \hline
    \end{tabular}
    \end{center}
    \end{minipage}
    \begin{minipage}[t]{0.35\linewidth}
    % \centering
    % \caption*{(b)}
    \begin{center}
    (b)~$3p_{1/2}$ manifold\\
    % \label{tab22}
    $M=0$\\
    \begin{tabular}{|c|cc|}
        \hline
        States & $\ket{i}_1$ & $\ket{i}_2$ \\
        \hline
        $\ket{i}_1$ & 0 & 0 \\
        %\hline
        $\ket{i}_2$ & 0 & -188.9 \\
        \hline
    \end{tabular} \vspace{5pt}\\
    $M= \pm 1$ \\
    \begin{tabular}{|c|cc|}
        \hline
        States & $\ket{i}_1$ & $\ket{i}_2$ \\
        \hline
        $\ket{i}_1$ & -47.22 & 81.79 \\
        %\hline
        $\ket{i}_2$ & 81.79 & -141.7 \\
        \hline
    \end{tabular} \vspace{5pt}\\
    $M= \pm 2$\\
    \begin{tabular}{|c|c|}
        \hline
        States & $\ket{i}_2$ \\
        \hline
        $\ket{i}_2$ & 0 \\
        %\hline
        \hline
    \end{tabular}
    \end{center}
    \end{minipage}
\end{table}
%-----------------------------Tab3-end------Na-1---------------

A remarkable property of the MS basis vectors built above and depicted in \figref{figSh} is a comparatively simple inclusion of the HF interaction at the first and last steps of the ladder excitation scheme.
The g-basis vectors of the first ground step is specially chosen as a set of eigenvectors $|g\rangle _{F''M}\!=\!|g,F''M \rangle $ of the HF operator, where it is diagonal.
Accordingly, taking into account the HF interaction is reduced, therefore, to tabular HF energy shifts of g-sublevels $|g \rangle _{\eta} $ depicted in Fig.~\ref{figSh}.

When dealing with $ns_{1/2}$ state at the third f-step, the formal mathematical description of the linkage diagram in \figref{figbreak} implies
 \begin{multline}
\hat{V}^{(S)}\hat{V}^{(P)}/ \Omega _{1/2}^{(S)}\Omega _{1/2}^{(P)}\cdot|g ,s_{1/2}m_{J}'' \rangle |m_{I}''\rangle = \\ |f ,s_{1/2}m_{J}\rangle |m_{I}\rangle \delta_{m_{J}m_{J}''}\delta_{m_{I}m_{I}''}
\label{Fns2}
\end{multline}
\noindent
It is seen, that the operator on the left-hand side of Eq.~(\ref{Fns}) preserves the fine structure basis vectors when mapping the subspace $\Lambda_g$ to $\Lambda_f$.
Therefore, $\hat{V}^{(S)}\hat{V}^{(P)}$ must also preserve the quantum indices $F''M$ of  HF basis vectors, i.e. map $|g, F''M\rangle$ to $|f, F=F''M\rangle$, how it is depicted in \figref{figSh}.
This fact corresponds to the two-photon selection rules $\Delta F=0$, which opens up perspectives for the selective excitation of the HF components, as it is discussed in Introduction.
The HF operator results in the conventional HF energy shifts $ \varepsilon _{HF}$ of the f-basis vectors $|f\rangle _{F''M}\!=\!|f,F=F''M \rangle $ without mixing them.

In the case where the last step subspace $\Lambda_f$ corresponds to $nd$ states, the HF relative energy shifts turn out to be rather feeble (less than $0.3$ MHz even for $n\!=\!4$ for Na atoms- see Table~\ref{tabB1} in App.~\ref{App1} and less than $2$ MHz for $n\!=\!10$ in Rb case) and may be ignored.

Importantly, the MS i-basis of subspaces $\Lambda_i$, related to the intermediate ladder step, does not, unfortunately, diagonalize the HF operator.
This is well seen from Tables~\ref{tab21_tab22}--\ref{tab23_tab24} which represent the matrix elements of the HF operator in MS i-basis's for Na and Rb atoms.
The diagonal elements of the tables give the shifts $ \varepsilon _{i}$ of the initial RWA state energies $ \varepsilon _{i}=0$ (see the notations in \figref{figSh}), while the off-diagonal elements are Rabi frequencies of mixing between different MS $i$-states induced by the HF interaction.

%Tab1-%beg--------Rb 1----------------
\begin{table}%[H]
    \caption{
    Matrix elements of the HF operator in the MS basis of $\Lambda_i$ subspaces (a)~$5p_{3/2}$ and (b)~$5p_{1/2}$ for manifolds with different $M$, in units of MHz for Rb atoms.
    }
    \label{tab23_tab24}
    \begin{minipage}[t]{0.60\linewidth}
    % \centering
    % \caption{}
    (a)~$5p_{3/2}$ manifold\\
    % \label{tab23}
    $M=0$\\
    \begin{tabular}{|c|cccc|}
        \hline
        States & $\ket{i}_1$ & $\ket{i}_2$ & $\ket{i_{0}}$ & $\ket{i_{1}}$ \\
        \hline
        $\ket{i}_1$ & -114.6 & 0 & 114.6 & 0 \\
        %\hline
        $\ket{i}_2$ & 0 & 224.3 & 0 & 127.1 \\
        %\hline
        $\ket{i_{0}}$ & 114.6 & 0 & -114.6 & 0 \\
        $\ket{i_{1}}$ & 0 & 127.1 & 0 & -114.6 \\
        \hline
    \end{tabular} \vspace{5pt}\\
    $M= \pm 1$ \\
    \begin{tabular}{|c|ccc|}
        \hline
        States & $\ket{i}_1$ & $\ket{i}_2$ & $\ket{i_{0}}$ \\
        \hline
        $\ket{i}_1$ & -98.1 & -34.0 & 68.0 \\
        %\hline
        $\ket{i}_2$ & -34.0 & 201.6 & 130.2 \\
        %\hline
        $\ket{i_{0}}$ & 68.0 & 130.2 & 6.25 \\
        \hline
    \end{tabular} \vspace{5pt}\\
    $M= \pm 2$\\
    \begin{tabular}{|c|cc|}
        \hline
        States & $\ket{i}_2$ & $\ket{i_{0}}$ \\
        \hline
        $\ket{i}_2$ & 133.3 & 133.3 \\
        %\hline
        $\ket{i_{0}}$ & 133.3 & 133.3 \\
        \hline
    \end{tabular}
    \end{minipage}
    \begin{minipage}[t]{0.35\linewidth}
    % \centering
    % \caption{}
    (b)~$5p_{1/2}$ manifold\\
    % \label{tab24}
    $M=0$\\
    \begin{tabular}{|c|cc|}
        \hline
        States & $\ket{i}_1$ & $\ket{i}_2$ \\
        \hline
        $\ket{i}_1$ & 0 & 0 \\
        %\hline
        $\ket{i}_2$ & 0 & -814.5 \\
        \hline
    \end{tabular} \vspace{5pt}\\
    $M= \pm 1$ \\
    \begin{tabular}{|c|cc|}
        \hline
        States & $\ket{i}_1$ & $\ket{i}_2$ \\
        \hline
        $\ket{i}_1$ & -203.6 & 352.7 \\
        %\hline
        $\ket{i}_2$ & 352.7 & -610.9 \\
        \hline
    \end{tabular} \vspace{5pt}\\
    $M= \pm 2$\\
    \begin{tabular}{|c|c|}
        \hline
        States & $\ket{i}_2$ \\
        \hline
        $\ket{i}_2$ & 0 \\
        %\hline
        \hline
    \end{tabular}
    \end{minipage}
\end{table}
%-----------Rb 1-------------Tab3-end-----------------------------

%section3
\section{\label{AT} Autler-Townes spectra}

The simple architecture of the excitation schemes we are dealing with in Fig.~\ref{figSh}, allows, within the framework of perturbation techniques for the HF interaction, an explicit description of the adiabatic states structure formed by a strong S-laser (in what follows, we will assume the probe P laser to be weak).
 We will show in the  subsection~\ref{Peaks} an interesting specificity of dressed states energies dependent on the reduced Raby frequencies that results in the formation of AT multiplet peaks structure.
Various aspects of dressed states manifestation (dark, bright, and chameleon peaks) will be demonstrated in the numerical modeling of the AT spectra in subsection~\ref{simulations} with focusing on the role of  the HF operator in the appearance of the main types of AT signals.
In particular, special attention (Subsec.~\ref{BrAndCh}) will be paid to the HF  mixing of bright and chameleon states shown in Fig.~\ref{figSh}.

\subsection{\label{Peaks} Formation of multiplet structure in AT signals}

In the zeroth approximation, the part of HF operator related to off-diagonal (mixing) terms of its matrix at the intermediate i-step of the ladder (see Tables~\ref{tab21_tab22} -- \ref{tab23_tab24}), can be neglected.
The field operator $\operator{V}^{(S)}$ is reduced to two-dimension matrices $\operator{M}^{(S)}_{\varsigma}$ acting within independent sets of two-level combinations ($\varsigma $-pairs) of bright $\ket{i,f}_{F''M}$ ($\varsigma \! =\! F''$) or chameleon $\ket{i,f_{\xi M}}$ ($\varsigma \! =\! \xi $) MS states with two MS effective Rabi frequencies $\Omega _{\varsigma=F''}^{eff}=\Omega _{1/2}^{(S)}$ or $\Omega _{\varsigma=\xi }^{eff}=\Omega _{3/2}^{(S)}$ respectively (see Fig.~\ref{figSh} and Eqs.~(\ref{mapping}), (\ref{Fns})).
The corresponding bare energies $\widetilde{\varepsilon} \! =\!\varepsilon \! +\! \varepsilon _{Hf}$ are determined by the initial RWA energies $\varepsilon $ along with the diagonal matrix elements $\varepsilon _{Hf}$ of the HF operator (see the discussion in Subs.~\ref{HFI}) which we, unlike the off-diagonal ones, explicitly take into account.

%----------------
%Fig4beg-----------------------------
\begin{figure}
    \centering
    \includegraphics[width=\linewidth]{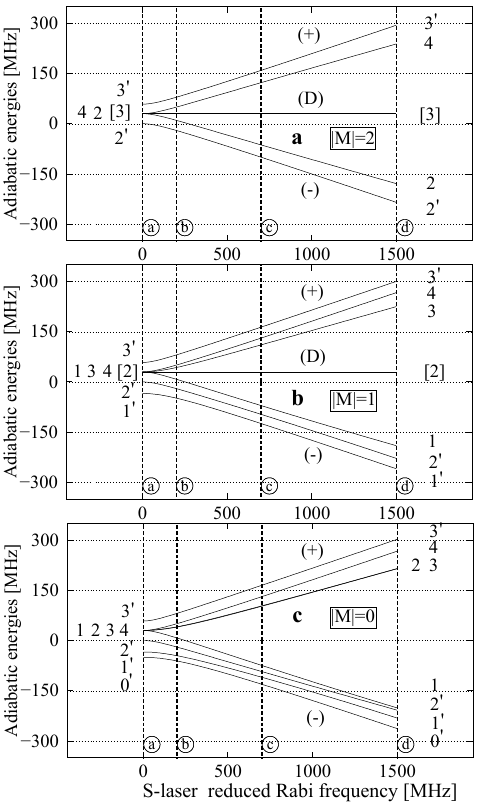}
    \caption{
Energies $\varepsilon $ of adiabatic (dressed) states in Na vs the S-laser reduced Rabi frequency $\Omega_{S}$~(\ref{rabi1}) for the $3s_{1/2}(F'',M)\!\to \!3p_{3/2}(F',M)\!\to \! 4d_{5/2}(F,M)$ excitation Zeeman sequences ($M\!=\!0,1,2$) in the case of pump laser detuning $\Delta_{S}\!=\!-30\,\text{MHz}$.
    The dashed vertical lines correspond to $\Omega_{S}$-values presented in Fig.~\ref{4D_chameleons} frames (a)-(d).
    The square brackets indicate the states that transfer into dark states.
    The zero-energy positions related to the energy of the $3p_{3/2}(F'\!=\!2)$-state.
    The curves labeled with quantum numbers $F'$ or $F$ refer to the dressed states starting to evolve at $\Omega_{S}=0$ from the corresponding HF basis vectors $\ket{\gamma ,F,M}$ ($\gamma\!=\!i,f$).
    }
    \label{4d_energy}
\end{figure}
%-----------------------------Fig4-end-----------------------------

The diagonalization of the matrix operator $\operator{M}^{(S)}_{\varsigma }$, coupling $\varsigma $-diabatic vector pair $\ket{i,f}$, results in ac Stark shift \cite{Shore2009,Delone_1999}
\begin{align}
    \label{eq:DressedEnergies}
    \ \varepsilon_{\varsigma\pm}\!=\!\frac{1}{2}\left[ (\widetilde{\varepsilon }_f\!+\!\widetilde{\varepsilon }_i) \!\pm\! \sqrt{\Delta{\widetilde{\varepsilon }}_{\varsigma}^{2}\!+\!(2\Omega_{\varsigma}^{eff})^{2}} \right]; \, \Delta{\widetilde{\varepsilon }}_{\varsigma}\!=\!\widetilde{\varepsilon }_f\!-\!\widetilde{\varepsilon }_i
\end{align}
\noindent of the pair.
The corresponding two eigenvectors \cite{Dalibard1985}
\begin{equation}
    \begin{split}
    & \ket{\varsigma+}=\cos \theta \ket{i} + \sin \theta \ket{f}; \\
    & \ket{\varsigma-}=-\sin \theta \ket{i} + \cos \theta \ket{f},
    \end{split}
    \label{eq:DressedState}
\end{equation}
\noindent belong to the set of $\varsigma $-pairs of repulsive zero-order adiabatic (dressed) states.
Here the mixing angle $2\theta=\arctan (-2\Omega _{\varsigma}^{eff}/\Delta{\widetilde{\varepsilon }}_{\varsigma})$ ($0\!\leq \! \theta \! \leq \! \pi /2$) provides a measure of amplitudes sharing between $\varsigma $ -diabatic vectors $\ket{i},\ket{f}$.

Strong coupling implies that both $\Omega_{1/2,3/2}^{(S)}$ exceed   the bare energy separations $\Delta{\widetilde{\varepsilon }}_{\varsigma}$~(\ref{eq:DressedEnergies}) for all coupling pairs, so that at large reduced frequency $\Omega _S$  (see its  definition in Appendix~\ref{App1}, Eq.~(\ref{rabi1})) relation~(\ref{eq:DressedEnergies}) reduces to the simple linear form:
\begin{equation}
    \varepsilon _{\varsigma \pm}\!=\! \pm \Pi _{\varsigma } \Omega_S+\overline{\varepsilon} _{\varsigma }; \quad \overline{\varepsilon} _{\varsigma }
    =\left (\widetilde{\varepsilon }_f+\widetilde{\varepsilon }_i \right )/2
    \label{slope}
\end{equation}
\noindent with the slope coefficients $\Pi_{\varsigma }\!=\!\Omega_{\varsigma}^{eff}/\Omega _S$ determined by the effective MS frequencies.
Actually, there are only two values $\Pi_{1/2,3/2}$ relate to bright ($\Omega ^{(S)}_{1/2}$) and chameleon ($\Omega ^{(S)}_{3/2}$) pairs.
Since a set of single Dark states is not involved in the interaction with laser fields, their energies $ \varepsilon _D$ should not depend on $\Omega _S$, i.e. formally $\Pi_{D}=0$.
Noteworthy, the presence of HF mixing between the MS basis vectors (see Tables~\ref{tab21_tab22}--\ref{tab23_tab24}) can insignificantly affect the values of both $\varepsilon _D$ and $\overline{\varepsilon} _{\varsigma }$ with variations of $\Omega _S$.

The above properties of dressed states energies are illustrated by the data exhibited in Figs.~\ref{4d_energy}, \ref{SelecEnerg_3/2}.
Two branches of ``repulsive'' energies $\varepsilon _{\varsigma \pm}$ are clearly visible, the upper (+) and the lower (-), as well as the horizontal (D), ``dark'' ($\Pi_{\varsigma }=0$) branches.
Noteworthy, due to the mirror symmetry $\sigma_{z}$ about any plane containing the quantization $z$-axis, the dressed states energies do not depend on the sign of $M$
\cite{Sobelman1992,Landau1981}.

%-----------------------------Fig5-beg-----------------------------
\begin{figure}
    %\centering
    \includegraphics[width=\linewidth]{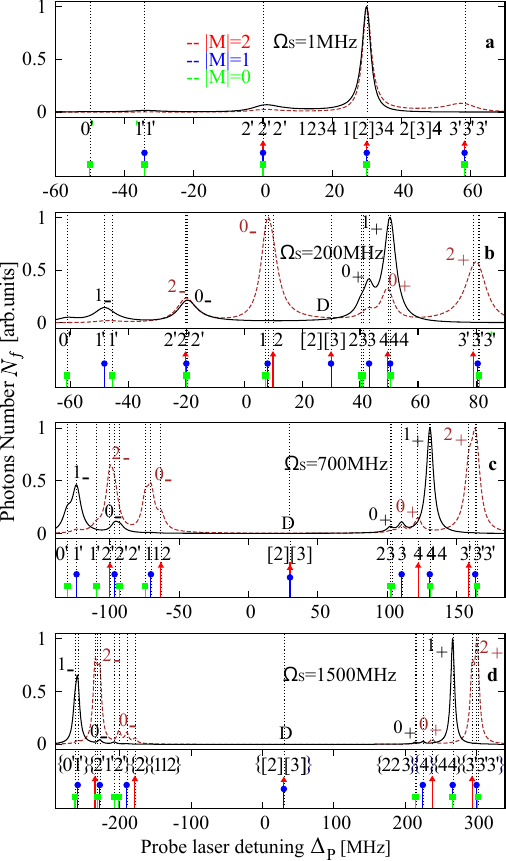}
    \caption{
    (Color online) The total number $N_f$ of photons emitted by the final $4d_{5/2}$ state of Na atoms vs probe field detuning for the $3s_{1/2}(F'')\to 3p_{3/2}\to 4d_{5/2}$ excitation sequences with $F''= 1$ (black solid curves) and $F''= 2$ (brown dashed curves).
    The vertical lines show the expected AT peaks (see in \figref{4d_energy}): the green squares indicate energies of adiabatic states with $M=0$, the blue circles-with $|M|=1$ and the red arrows-with $|M|=2$.
    If several expected lines partially merge and form a complex multiplet at $\Omega_{S}=1500\, \text{MHz}$ they are incorporated with braces into a block of numbers.
    The blocks refer to resolved AT components which are marked with symbols $N_{\pm}$ as defined in subsection~\ref{Peaks}.
    The ``D'' symbol indicates ``Dark'' components.
    }
    \label{4D_chameleons}
\end{figure}
%-----------------------------Fig5-end--

Each $\varsigma$-pair of dressed states (\ref{eq:DressedState}) results in the formation of a peaks pair ($\pm $) in AT spectra, diverging with an increase of the Rabi frequency $\Omega _S$, which is well observed in Figs.~\ref{4D_chameleons}, \ref{Selective_3/2} (see a more detailed discussion in the next subsection).
Different diverging AT peaks, corresponding to identical slope coefficients $\pm \Pi _{1/2}$ (for bright peaks) or $\pm \Pi _{3/2}$ (for chameleon peaks) in Eq.~(\ref{slope}), may partially merge at large $\Omega _S$, forming complex multiplets (see \figref{4D_chameleons}).
It is convenient to label the ($ \pm $)-components of those multiplets by the symbols ${N}_{\pm}$, where the integer ${N}$ takes values ${N} \! =\! F''\! = \! 1,2$ for bright $(\varsigma \! =\! F'')$ and $N\! =\! 0$ for chameleons ($\varsigma \! =\! \xi$) $\varsigma $-pairs.
In the case of Figs.~\ref{4d_energy}, \ref{4D_chameleons}, two values of $\Pi _{\varsigma}$, namely $\Pi _{1/2}$ for ($1_{\pm} ,2_{\pm}$)- and $\Pi _{3/2}$ for ($0_{\pm}$)-multiplets, can be distinguished.
Their ratio turns out to be $\Pi _{1/2}/\Pi _{3/2}\!=\!\sqrt{1.5}$ in accordance with Eqs.~(\ref{RabiEfS}), (\ref{rabiFS}).
In the figures, we also denote multiplets related to dark states with the symbol ``D''. In what follows, the multiplet symbols are combined into one ``$ \Upsilon $'', which runs through the meanings $ \Upsilon = 1_ {\pm}, 2_ {\pm}$ for bright- ($ \Upsilon _{Br}$), $0_ {\pm}$ for chameleon- ($ \Upsilon _{Ch}$), and $D $ ($ \Upsilon _{D}$) for dark-multiplets.
The block of dressed states involved in the formation of a multiplet $ \Upsilon $ will be termed a ``$ \Upsilon $''-block.

A semiquantitative account of the effects of hyperfine interaction on AT spectra is possible within the framework of perturbation theory under assumptions that (i) the splitting between the components of the multiplets is small and (ii) the HF mixing rate between the dressed states, incorporated in two different blocks $ \Upsilon $ and $ \Upsilon '$, can be replaced by one effective Rabi frequency $\Omega _{\Upsilon ,\Upsilon '}^{(HF)}$.
The latter is the average value $\left \langle \Omega _{HF} \right \rangle$ for the corresponding frequencies presented in Tables~\ref{tab21_tab22} -- \ref{tab23_tab24}.

 Without HF interaction, the probe laser is capable of exciting only one bright $ N$-pair (for instance, $N\!=\!2$ in the case $F''\!=\!2$), resulting in the appearance of two multiplets $ \Upsilon _{Br} \!=\!N_{\pm} $.
The visualisation of another multiplet $ \Upsilon '$ ($\!=\!1,0$) occurs due to the HF mixing of states from blocks $ \Upsilon '$ and $ \Upsilon _{Br}$.
The corresponding ratio
\begin{align}
    \label{HFaccounting}
    I_{\Upsilon '}/I_{\Upsilon _{Br}}\approx R=\frac{\left (\Omega _{\Upsilon ',\Upsilon _{Br}}^{(HF)} \right )^2}{|\Delta \varepsilon _{\Upsilon ',\Upsilon _{Br}}|^2 }
\end{align}
\noindent of the multiplets intensities can be estimated as the factor $ R $
that determines the distribution of the population between the blocks with the energy shift $\Delta \varepsilon _{\Upsilon ',\Upsilon _{br}}$ and the rate coupling $\Omega _{\Upsilon ',\Upsilon _{Br}}^{(HF)}$ \cite{Landau1981,Kirova2017}.

The numerical experiments presented below detail our theoretical findings.

%-----------------------------Fig6-beg-----------------------------
\begin{figure}
    %\centering
    \includegraphics[width=\linewidth]{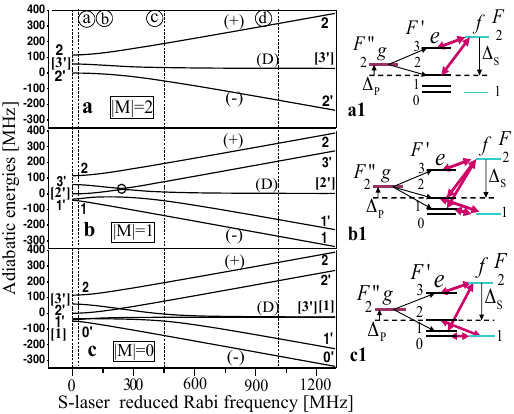}
    \caption{
    (Color online) The same as in \figref{4d_energy} in the case of $3s_{1/2}(F''=2)\to 3p_{3/2}\to 5s_{1/2}$ excitation sequence in Na for different $S$-laser reduced Rabi frequencies $\Omega _S$.
    The corresponding RWA linkage diagrams are shown in frames (a1), (b1), (c1).
    Simulations are performed using parameters $\Omega_{P}=0.87\,\text{MHz}$, $\Delta_{S}=-115\,\text{MHz}$.
    }
    \label{SelecEnerg_3/2}
\end{figure}
%-----------------------------Fig6-end----------------------------

\subsection{\label{simulations} Numerical simulation }

The Autler-Townes spectra are an important source of information about the atomic systems under study.
Here we simulate the Doppler-Free experimental conditions of work \cite{Bruvelis2012} where a supersonic beam of  Na atoms, having the mean flow velocity of 1160 m/s, is excited by two counterpropagating S- and P-laser beams of the same linear polarizations.
The amplitudes $E^{(S,P)}$ space distribution of the laser electrical fields corresponds to Gaussian switching of the Rabi laser frequencies (\ref{rabi1}): $\Omega _{S,P}\! \rightarrow \!\Omega _{S,P}\exp (-2t^2/\tau _{S,P}^{2})$ with characteristic times $\tau _{S}$=1050 ns and $\tau _{P}$=350 ns of flight of atoms through the S- and P-laser beams.
The simulations with Rb atoms correspond to Doppler-free experiments with cold atoms in optical dipole traps \cite{Ryabtsev2016}, while the time dependences of lasers Rabi frequencies $\Omega _{S,P}$ are identical to those for Na atoms.

In our numerical simulations (the corresponding algorithm is described in Appendix~\ref{App1}), we studied the temporal evolution of the atomic density matrix $\rho_{\gamma FM, \widetilde{\gamma} \widetilde{F} \widetilde{M}}(t)$ when the zone of laser interaction is crossed by a single unexcited atom that has only one populated HF g-component $F''$ with the equilibrium population $n_g(M'')\!=1\!/(2F''\!+\!1)$ of the corresponding Zeeman sublevels.

%-----------------------------Fig7-beg-----------------------------
\begin{figure}
    %\centering
    \includegraphics[width=\linewidth]{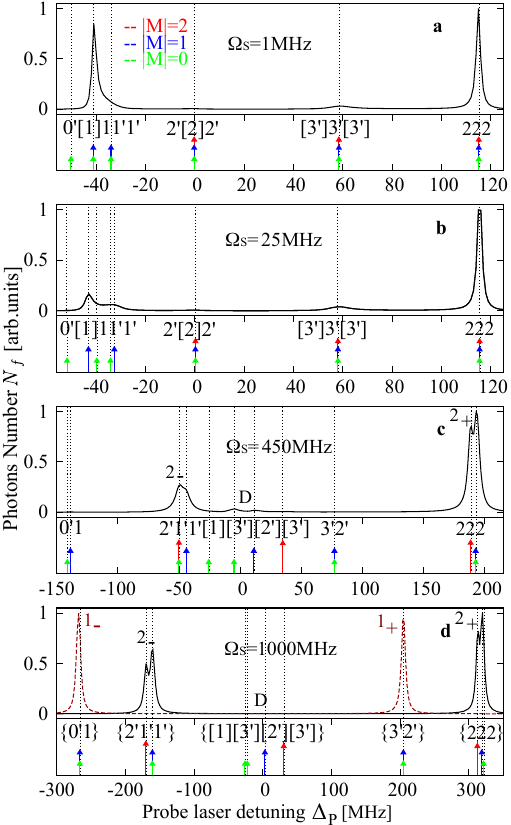}
    \caption{
    (Color online) The same as in \figref{4D_chameleons} in the case of $3s_{1/2}(F''=2)\to 3p_{3/2}\to 5s_{1/2}$ excitation sequence for different $S$-laser reduced Rabi frequencies $\Omega _S$.
    Simulations are performed for Na atoms using parameters $\Omega_{P}=0.87\,\text{MHz}$, $\Delta_{S}=-115\,\text{MHz}$.
    The expected AT peaks (the vertical lines) are consistent with the data in \figref{SelecEnerg_3/2}.
    }
    \label{Selective_3/2}
\end{figure}
%-----------------------------Fig7-end---------------------------

\subsection{\label{BrAndCh} ``Bright'' and ``Chameleons'' multiplets }

Typical AT spectrum for Na atoms, used as an example in this section, are exhibited in \figref{4D_chameleons} where one can observe ``bright'' $\Upsilon _{Br}$ ($1_{\pm},2_{\pm}$) and ``chameleons'' $\Upsilon _{Ch}$ ($0_{\pm}$) multiplets.
The characteristic feature of these two types of AT peaks, which makes them related, is increasing separation between their ($\pm$)-brunches with increasing $\Omega _S$.

However, the chameleon $0_{\pm}$-components arise due to the HF interaction that shares the population between the bright $\ket{i}_{F''M}$ and chameleon $\ket{i _{\xi ,M}}$ diabatic states enabling thus the latter's to be excited from the ground state with a probe laser (see Table~\ref{tab21_tab22}(a)).

The intensities $I_{\Upsilon _{Br}}$ of the chameleons multiplets are evaluated by Eq.~(\ref{HFaccounting}), where the values of the mixing rates $ \Omega _{\Upsilon _{Ch},\Upsilon _{Br}}^{(HF)} \! \simeq \! \left \langle \Omega _{HF} \right \rangle$ do not exceed $30$ MHz as follows from the data in Table~\ref{tab21_tab22}(a).
Importantly, the frequency/energy shifts $\Delta \varepsilon _{\Upsilon _{Ch},\Upsilon _{Br}}$ between the Bright and Chameleon AT multiplets behave as $\Delta \varepsilon _{1/2,3/2}\sim (\Pi _{3/2 }-\Pi _{1/2 })\Omega _S$ in accordance to Eq.~(\ref{slope}).
Consequently, the $0_{\pm}$-peaks associated with the chameleon states should fade with increasing $\Omega _S$, which is confirmed by the graphs in \figref{4D_chameleons}.
The latter feature is a characteristic property of dark states, which is discussed below.

A more detailed study of the chameleon states is presented in \cite{Cinins2021}, where their two main types are defined, namely ``slow'' and ``fast''.

\subsection{\label{Dark} ``Dark'' multiplets }

As can be seen from Figs.~\ref{figSh} and \ref{4d_energy}, all dark states related to the $3p_{3/2}\to 4d_{5/2}$ transitions lie in the final $\Lambda _f$ subspace, where, in the case of Na atoms, they have negligible HF mixing frequencies $\Omega _{HF}$ with both bright and chameleon states.
For this reason, the corresponding dark multiplets are not visible according to Eq.~(\ref{HFaccounting}).
Another excitation scheme $3s_{1/2}(F''=2)\to 3p_{3/2}\to 5s_{1/2}$ is more convenient for observing the dark states since all of them are situated in the intermediate $\Lambda _i$ subspace (see Table~\ref{tab00} and \figref{SelecEnerg_3/2}).

Importantly, formally the Maris-Shore frequencies $\Omega _{3/2}^{(S)}$ (\ref{RabiEfS}) turn out to be equal to zero, so that the states $\ket{i_{\xi}}$ shown in \figref{figSh} and referred to above as ``Chameleons'' change their status (``colour'') to ``Dark''.
The HF mixing rates $\Omega _{HF}$ in Table~\ref{tab21_tab22}(a) between the bright and now dark states reach values ~$30$ MHz, which makes it possible to observe the dark multiplets ``D'', provided that their shifts $\Delta \varepsilon _{Br,D}$ from the bright multiplets $1_{\pm},2_{\pm}$ do not exceed these $30$ MHz.
It is worth bearing in mind, that since the dark states are excluded from the laser-atom interaction, their energies are not significantly changed upon growth of $\Omega _S$ in contrast to the bright states, which quickly run away to the left and right spectral edges (see \figref{SelecEnerg_3/2}) with a simultaneous increase in the shifts $\Delta \varepsilon _{Br,D}\sim \Pi _{1/2 }\cdot \Omega _S$.
As a result, the dark multiplets are washed out from the AT signals for large $\Omega _S $.

Our theoretical predictions are illustrated by the data presented in Figs.~\ref{SelecEnerg_3/2}, \ref{Selective_3/2}.
Figure~\ref{SelecEnerg_3/2} depicts (frames (a1)-(c1)) the linkage diagrams associated with the $3p_{3/2}\to 5s_{1/2}$ couplings by S-laser for three possible Zeeman numbers $|M|=0,1,2$, while \figref{Selective_3/2} gives AT spectra for several values of the Rabi frequency $\Omega_{S}$.
Figure \ref{SelecEnerg_3/2} exhibits also the calculated dressed states energy diagrams versus $\Omega_{S}$, which are then used in \figref{Selective_3/2} to indicate the expected positions of the AT peaks.
It is clearly seen that dark multiples disappear with increasing S-laser intensity.

%section4
\section{\label{selectivity} Control of selective excitation of unresolved HF components}

Three features of the discussed results of numerical simulations should be noted.

(i) The initial AT spectrum at very low $\Omega_{S}$ demonstrate a typical two-step excitation pattern with one (\figref{4D_chameleons}) or two (Figs.~\ref{Selective_3/2}, \ref{UnSelective_1/2}) peaks due to two-photon resonances arising when the RWA energy $\varepsilon _g$ of the ground wave vector $\ket{F''M}$ coincides with that $\varepsilon _f$ for any HF sublevels $\ket{FM}$ of the final state.

(ii) The value $\Omega _{S}$ introduced by Eq.~(\ref{rabi1}) is a rather formal frequency associated with optical transitions between intermediate and final quantum states $\ket{nl}$.
Both fine and hyperfine interactions significantly diminish the partial Rabi frequencies (\ref{rabi2}) along with the MSF $\Omega ^{(S)}_{1/2,3/2}$ (\ref{RabiEfS}), which are actually involved in the coupling of diabatic wave vectors and the formation of repulsive adiabatic states pairs (\ref{eq:DressedEnergies}).
It can be seen from \figref{4D_chameleons} that, for example, the value $\Omega _{S}\!=\!1500$ MHz corresponds to $2\Omega _{1/2}^{(S)} \!\simeq \!550$ MHz, which is equal to the frequency shift between the two bright sidebands $2_{\pm}$ (see Eq.~(\ref{eq:DressedEnergies})).
In the case of subsequent \figref{UnSelective_1/2}, $\Omega _S\!=\!2450$ MHz corresponds to $2\Omega ^{(S)}_{1/2} \!\simeq \! 820$ MHz, which is noticeably less than the HF splitting of $1772$ MHz in Na.
A similar situation occurs for Rb atoms, where the reduced Rabi frequency $\Omega _{CL}=4000$MHz of the auxiliary control laser (see x-axis of \figref{selectivity_RbNa}) results in a value of $ \sim 1500/2$ MHz for the corresponding actual MSF $\Omega ^{(CL)}_{1/2}$.
For this reason, the control laser is unable to couple both ground HF components at the same time.

\subsection{\label{GenSelect} Spectral composition and selection rules}

%-----------------------------Fig 8-beg-----------------------------
\begin{figure}
    %\centering
    \includegraphics[width=\linewidth]{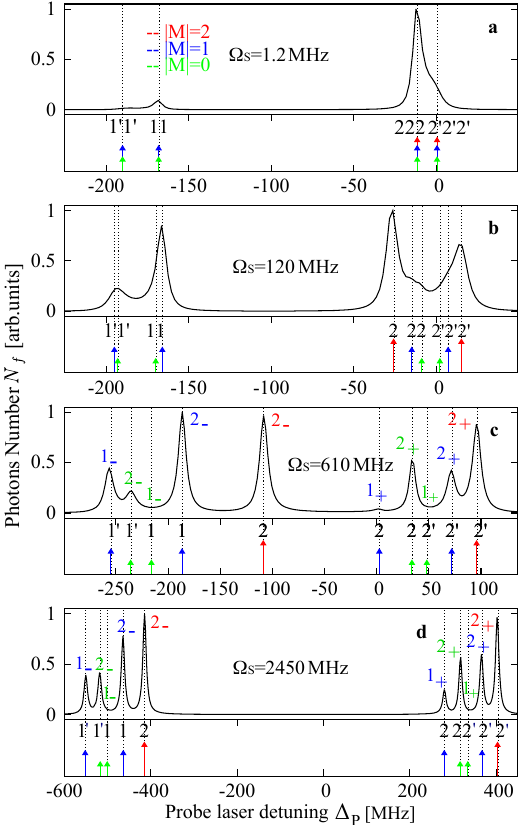}
    \caption{
    (Color online) The same as in \figref{4D_chameleons} in the case of $3s_{1/2}(F''=2)\to 3p_{1/2}\to 5s_{1/2}$ excitation sequence for Na atoms with parameters $\Omega_{P}=1.22\,\text{MHz}$, $\Delta_{S}=12\,\text{MHz}$.
    Vertical lines show the expected AT peaks, obtained from calculated dressed states energies.
    }
    \label{UnSelective_1/2}
\end{figure}
%-----------------------------Fig8)-end-----------------------------

Before proceeding to the analysis of the third, interesting, and important from the practical point of view, feature of the discussed AT spectra, let us consider one more excitation scheme $3(5)s_{1/2}(F''=2)\to 3(5)p_{1/2}\to ns_{1/2}$.
In this sequence, the configurations of the corresponding linkage diagrams in \figref{figbreak} imply the absence of dark states (see also Tables~\ref{tab00}, \ref{tab21_tab22}(b), \ref{tab23_tab24}(b)), so that only ``bright'' peaks $1_{\pm},2_{\pm}$ should be observed.
The data in \figref{UnSelective_1/2} confirm that expectation.
The crucial difference from the two previous cases (see Figs.~\ref{4D_chameleons}, \ref{Selective_3/2}) at large coupling laser intensities lies in the noticeable resolution of all possible 8 singlet AT peaks, resulting from excitation of 8 (see below) dressed states.

\begin{figure}
    %\centering
    \includegraphics[width=\linewidth]{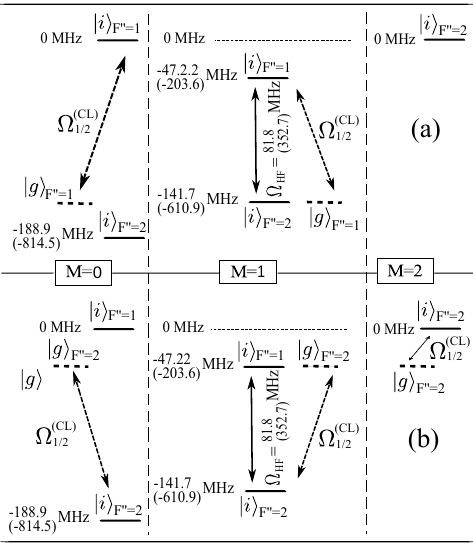}
    \caption{
    Energy levels diagrams (in MHz) for the states $\ket{i}_{F''M}$ from the Morris-Shore basis's in $\Lambda _i$ subspace $3p_{1/2}$.
    The HF energies of the MS $i$-states along with HF mixing frequencies $\Omega _{HF}$ are presented in Tables~\ref{tab21_tab22}(b), \ref{tab23_tab24}(b) for Na and Rb atoms.
    The dashed lines depict the RWA positions of the ground sublevels $\ket{g}_{F'',M}$ when the auxiliary control laser, (a) tuned to the transition $3(5)s_{1/2}(F''\!=\!1)\!\to\! 3(5)p_{1/2}$, has the detuning $\Delta_{CL}\!=\!-141.7(-610.9)\,\text{MHz}$; (b) tuned to the transition $3(5)s_{1/2}(F''\!=\!2)\!\to\! 3(5)p_{1/2}$, has the detuning $\Delta_{CL}\!=\!-47.22(-203.6)\,\text{MHz}$ (see Tab.~\ref{tab004}).
    }
    \label{control}
\end{figure}

Such an essential AT spectra beneficiation is amazing in view of a rather poor structure of the HF sublevels of the intermediate $ 3(5)p_{1/2}$ states and is explained by the relatively large ($ 189(816)$ MHz) HF splitting.
Figure~\ref{control} reproduces the data of Tables~\ref{tab21_tab22}(b), \ref{tab23_tab24}(b) in graphical form.
It is seen that three bright states (BS) pairs, namely $\ket{i,f}_{F''=2,M=0,1,2}$, can be directly excited by P-laser upon its probing the transition $ 3(5)s_{1/2}(F''\!=\!2)\!\to\! 3(5)p_{1/2}$.
As a result, the AT spectra should contain six bright $\Upsilon_{Br}$ picks $2_{\pm}(M\!=\!0,1,2)$, which form six well-resolved singlets due to significant HF shifts.
Noteworthy, all three final BS are elements of the HF basis $\ket{f}_{F=2,M}\!=\!\ket{F\!=\!2,M}$, since, as it was shown in subsection~\ref{HFI} the elementary excitation sequences~(\ref{mapping}) provide a specific two-photon selection rule $\Delta F\!=\!0$ in the case of $ns_{1/2}$ final states.

%-----------------------------Fig9-beg-----------------------------
%-----------------------------Fig9-end----------------------------

An essential HF coupling between BS's ($M\!=\!1$) may violate, however, the selective excitation of HF components with $F\!=\!2$.
Note that because all $\Upsilon_{Br}$ have the same slope coefficient $\Pi _{1/2}$ (\ref{slope}), their HF shifts $\Delta \varepsilon _{\zeta _ {\pm}M,\zeta '_{\pm}M'}$ (where $ \zeta \!=\!F''\!=\!1,2$) stabilize at large $\Omega _S$ at the values of $50 \div 150$ MHz (see \figref{UnSelective_1/2}) in accordance with the HF energy shifts $\Delta \varepsilon _{\zeta M,\zeta 'M'}$ of the corresponding BS $\ket{i}_{\zeta ,M}$ depicted in \figref{control}.
In particular, this shift of $94.5$ MHz for $\ket{i}_{\zeta =1,M=1}$ and $\ket{i}_{\zeta =2,M=1}$ BS turns out to be close to the HF mixing rate $ \Omega _{HF}= 81.8$ MHz, i.e. HF interaction leads to a strong population sharing between these BS's with the factor $R(M\!=\!1) \!\sim\! 1$ in Eq.~(\ref{HFaccounting}).

This means that the excitation of $2_{+}$ (or $2_{-}$) ($M\!=\!1$) singlet must be accompanied by other additional $1_{+}$ (or $1_{-}$) ($M\!=\!1$) one and vice versa.
The corresponding dressed sate is a mixture of two zero-order adiabatic states $\ket{\zeta=2+},\ket{\zeta=1+}($M\!=\!1$) $ (\ref{eq:DressedState}) of approximately equal weights and, hence, equal populations of two HF sublevels $F\!=\!1,2;M\!=\!1$ of the final state.
Figure~\ref{selectivity_1_2}(a) illustrates well the composite structure of the relevant fluorescent signal (see the further discussion in the next subsection).
Note that in the case of $M\!=\!0$ the HF coupling of BS's is absent.
The latter results in the preservation of the two-photon selection rule $\Delta F\!=\!0$ when two singlets $1_{+},1_{-}(M\!=\!0)$ become invisible in \figref{UnSelective_1/2}.

(iii) Now we can formulate the third important feature of the AT spectra: the depletion of the AT spectra, which manifests itself in (a) the merging of its components into multiplets with (b) the simultaneous absence of some bright peaks, makes it possible to ignore the HF mixing between bright states.

With regard to the spectra of Figs.~\ref{4D_chameleons}, \ref{Selective_3/2}, one cleanly sees both signs of spectra pauperisation, namely: (a) multiplet structure and (b) lack of all AT multiplets $1_{\pm}$ related to the excitation scheme $3s_{1/2}(F''\!=\!1)\rightarrow 3p_{3/2}$.
Bright multiplets $2_{\pm}$ and $1_{\pm}$ turn out to complement each other, i.e. their visualization occurs during independent probing from different initial HF components $F''\!=\!2,1$ of the ground state.
The data in Table~\ref{tab21_tab22}(a) do show that HF interaction poorly couples all $\ket{i}_1$ and $\ket{i}_2$ BS: the corresponding $\Omega _{HF}$ rates are $\Omega _{HF}\!=\!0$ for $M\!=\!0,2$ while for two BS with $M\!=\!1$ the sharing parameter~(\ref{HFaccounting}) $R\!=\!(7.44/65.6)^2\!=\!0.0129$.
Practically the same thing happens in the case of Rb atoms, as can be seen from the corresponding data in Table~\ref{tab23_tab24}(a): for both $M\!=\!0,2$ the values of $\Omega _{HF}\!=\!0$ and the the sharing parameter~(\ref{HFaccounting}) $R\!=\!(34.0/299.7)^2\!=\!0.0128$  for two BS with $M\!=\!1$.

Bypassing the effects of HF mixing leads to an important result: the proper choice of the initial wave vector $\ket{gF''M}$ corresponds to the address excitation of one of the independent (mutually orthogonal) blocks of three-level sequences~(\ref{mapping}).
This finding remains valid for arbitrary P-, S-laser intensities, and allows, as an example, STIRAP unitary transfer \cite{Vitanov2017} of initial populations from the ground state to the desired target Rydberg states (see details below).
Another application is related to the formation of an independent set of polaritons \cite{Fleischhauer2005,Maxwell2013,2016} involving different three-level sequences~(\ref{mapping}) as their carriers.

%-----------------------------Fig10-beg-----------------------------
\begin{figure}
    %\centering
    \includegraphics[width=\linewidth]{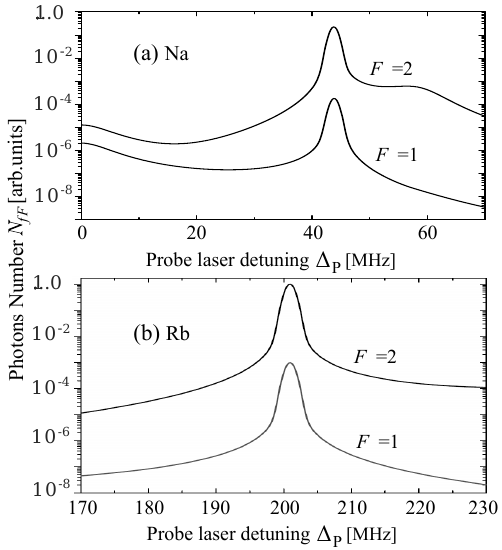}
    \caption{
    The total numbers $N_{fF}$ of photons emitted by the hyperfine components $F\!=\!1,2$ of the final $ns_{1/2}$ Rydberg state vs the probe field detuning for the $ks_{1/2}(F''=2)\to kp_{3/2}\to ns_{1/2}$ excitation sequences with the following parameters.
    (a) The case of Na atoms: $k=3, n=12$, $\Omega_{P}=0.1\,\text{MHz}$, $\Delta_{S}=-44.09\,\text{MHz}$, $\Omega_{S}=10\,\text{MHz}$.
    At resonance ($\Delta_{P} \simeq 44\,\text{MHz}$), infidelity factor (\ref{SelC}) $\Re _2 = 7.8\cdot 10^{-4}$.
    (b) The case of Rb atoms: $k=5, n=14$, $\Omega_{P}=0.1\,\text{MHz}$, $\Delta_{S}=-201.6\,\text{MHz}$, $\Omega_{S}=10\,\text{MHz}$.
    At resonance ($\Delta_{P} \simeq 202\,\text{MHz}$), infidelity factor (\ref{SelC}) $\Re _2 = 9.7\cdot 10^{-4}$.
    }
    \label{selectivity_3_2}
\end{figure}
%-----------------------------Fig10-end------------------------

\subsection{\label{ManipulationsC} Manipulations of HF interaction effects}

Importantly, when $ns_{1/2}$ states are used as final states, the sequences~(\ref{mapping}) implement a specific two-photon selection rule $\Delta F\!=\!0$, with the possibility of selective excitation of unresolved HF components of the final Rydberg states.
Modern applied problems require accurate information about the quantum numbers of the states under study.
In this subsection, we will consider some ways to eliminate unwanted HF interaction effects (in the case of excitation sequences with $M\!=\!\pm 1$) that reduce the precision of selective excitation.

%-----------------------------Fig11_Na-beg-----------------------------
\begin{figure}
    %\centering
    \includegraphics[width=\linewidth]{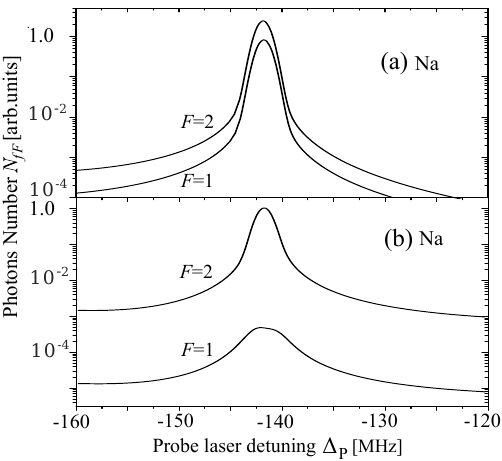}
    \caption{
    The same as in \figref{selectivity_3_2}(a) for the $3s_{1/2}(F''\!=\!2)\to 3p_{1/2}\to 12s_{1/2}$ excitation sequences in Na with parameters $\Omega_{P}\!=\!0.1\,\text{MHz}$, $\Omega_{S}\!=\!5\,\text{MHz}$, $\Delta_{S}\!=\!141.7\,\text{MHz}$.
    The auxiliary control laser has the detuning $\Delta_{CL}\!=\!-141.7\,\text{MHz}$ (see \figref{control}) with the following Rabi frequencies~(\ref{rabi1}): $\Omega_{CL}\!=\!0\,\text{MHz}$ (frame (a)) and $\Omega_{CL}\!=\!300\,\text{MHz}$ (frame (b)).
    At resonance ($\Delta_{P}\!\simeq \!-142\,\text{MHz}$), infidelity factor (\ref{SelC}) $\Re _2\!=\!0.24$ (frame (a)) and $\Re _2\!=\!4.7\cdot 10^{-4}$ (frame (b)).
    }
    \label{selectivity_1_2}
\end{figure}
%-----------------------------Fig11_Na-end---------------

Since we are concerned with Rydberg states which have small oscillator strengths ($\sim n^{-3}$ \cite{Sobelman1992}), the reduced Rabi frequencies $\Omega _S$ should be confined within the value of $\simeq 10$ MHz \cite{Lw2012}.
In modeling the AT spectra, we choose the $12s_{1/2}$ state in Na and $14s_{1/2}$ state in Rb as the final Rydberg states with zero HF splitting between its HF components.

%-----------------------------Tab4-beg-----------------------------
\begin{table}%[H]
    \caption{The $\Delta_{S,CL},\Delta_{P}^{(Re)}$ detunings used in our calculations for strong, control and probe lasers (see Fig.~\ref{selectivity_3_2}--\ref{selectivity_RbNa}) upon selective excitation of the HF $F$-component of Rydberg states $12( 14)s_{1/2}$, in units of MHz.}
    \label{tab004}
    \begin{tabular}{|c|c|c|c|c|}
    \hline
    $i$-state $kp_{J}$ & $kp_{3/2}$ & $kp_{3/2}$ & $kp_{1/2}$ & $kp_{1/2}$
    \\
    \hline
    HF component & $F=2$ & $F=1$ & $F=2$ & $F=1$ \\
    \hline
    Na, $\Delta _S$ & -44.09 & 21.47 & 141.7 & 47.22\\
    %\hline
    \; \, Na, $\Delta_{P}^{(Re)}$ & 44.09 & -21.47 & -141.7 & -47.22\\
    %\hline
    \; Na, $\Delta _{CL}$ & absent & absent & -141.7 & -47.22\\
    \hline
    Rb, $\Delta _S$ & -201.6 & 98.1 & 610.9 & 203.6\\
    %\hline
    \; \, Rb, $\Delta_{P}^{(Re)}$ & 201.6 & -98.1 & -610.9 & -203.6\\
    %\hline
    \; Rb, $\Delta _{CL}$ & absent & absent & -610.9 & -203.6\\
    \hline
    \end{tabular}
\end{table}
%-----------------------------Tab4-end-----------------------------

\begin{figure}
    %\centering
    \includegraphics[width=\linewidth]{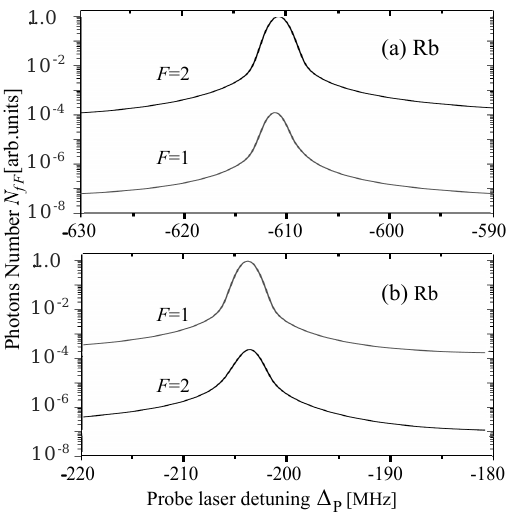}
    \caption{
    The same as in \figref{selectivity_3_2} for the $5s_{1/2}(F'')\!\to \!5p_{1/2}\!\to \!14s_{1/2}$ excitation sequences in Rb with the following parameters.
    Frame (a). $F''\!=\!2 $, $\Omega_{P}\!=\!0.1\,\text{MHz}$, $\Delta_{S}\!=\!610.9\,\text{MHz}$, $\Omega_{S}\!=\!5\,\text{MHz}$, $\Delta_{CL}\!=\!-610.9\,\text{MHz}$, $\Omega_{CL}\!=\!700\,\text{MHz}$.
    At resonance ($\Delta_{P}\!\simeq \!-611\,\text{MHz}$), infidelity factor (\ref{SelC}) $\Re _2 \!=\! 1.30\cdot 10^{-4}$.
    Frame (b). $F''\!=\!1 $, $\Omega_{P}\!=\!0.1\,\text{MHz}$, $\Delta_{S}\!=\!203.6\,\text{MHz}$, $\Omega_{S}\!=\!5\,\text{MHz}$, $\Delta_{CL}\!=\!-203.6\,\text{MHz}$, $\Omega_{CL}\!=\!700\,\text{MHz}$.
    At resonance ($\Delta_{P}\!\simeq \!-203\,\text{MHz}$), infidelity factor (\ref{SelC}) $\Re _1 \!=\! 2.4\cdot 10^{-4}$.
    }
    \label{selectivity_Rb_Rb}
\end{figure}

To enhance the population of the Rydberg states $ns_{1/2}$ with the chosen values of quantum numbers $F,M$ under moderate values of $\Omega _S$ and the suppressed influence of the HF interaction effects (see below), the S-laser must be tuned to a single-photon resonance with the transition $\ket{i} _{F''=F,M}(J) \!\to\! \ket{ns_{1/2}}$ involving a bright MS $i$-state $\ket{i} _{F,M}(J) $ at the intermediate $i$-step $kp_{J}$ of the ladder excitation.
The required $i$-states energy values $\varepsilon _i$ are on the diagonals of the matrices in the Tables~\ref{tab21_tab22}--\ref{tab23_tab24} (frames $M=\pm 1$), and the above S-laser resonance tunings correspond to its detunings $\Delta _S\!=\!-\varepsilon _i$ presented in Tab.~\ref{tab004}.
The strongest peak in the AT spectrum will arise when the probe P-laser scans the frequency region in the vicinity of its detuning $\Delta _P^{(Re)}\! = \!\varepsilon _i$, responsible for the two-photon resonance (see Tab.~\ref{tab004} and Figs.~\ref{selectivity_3_2}-- \ref{selectivity_Rb_Rb}).

The accuracy of excitation selectivity can be judged from deviation from the ideal $100\%$ selectivity using infidelity factors $\Re$, defined as the ratio
\begin{equation}
    \Re _2\!=\!\frac{N_{fF=1}}{N_{fF=1}\!+\!N_{fF=2}};
    \quad
    \Re _1\!=\frac{N_{fF=2}}{N_{fF=1}\!+\!N_{fF=2}}
    \label{SelC}
\end{equation}
\noindent of the partial AT signals $N_{fF}$ corresponding to the total number of photons~(\ref{ATsygnal}) emitted by each Rydberg HF sublevel $F\!=\!1$ or $F\!=\!2$.
The first relation in the equation~(\ref{SelC}) is used when we apply the sequence $ks_{1/2}(F'')\to kp_{1/2}\to ns_{1/2}$ with $F ''\!=\!2$ for selective excitation of  HF sublevels $F=2$ of $ns_{1/2}$ states.
The second relation is applied in the case of selective excitation of sublevels $F=1$, realized at $F ''\!=\!1$.

%-----------------------------Fig12_Rb_F=2-beg-----------------------------
%-----------------------------Fig12_Rb_F=2-end---------------

The coefficients $\Re$ are determined by the sharing factor $R$~(\ref{HFaccounting}), accounting for population transfer between different bright $i$-states due to HF interaction.
Small values of $R$ result in higher degree of selectivity.
This statement is illustrated by the data in \figref{selectivity_3_2} where $\Re _2$ (\ref{SelC}) reaches the value $7.8\cdot 10^{-4}$ for Na atoms and $\Re _2 =9.7\cdot 10^{-4}$ for Rb atoms in the vicinity of two-photon resonances $\Delta_{P}\!\simeq \!44.1\,\text{MHz}$ for Na and $\Delta_{P}\!\simeq\!202\,\text{MHz}$ for Rb.
Noteworthy, although the HF splittings of the Na and Rb atoms differ significantly (by a factor of $\sim 4.5$), their infidelity factors values turn out to be close due to almost identical $R$-factor values.

%----------------------------Fig13_RbNa_F=1,2-beg-----------------------------
\begin{figure}
    %\centering
    \includegraphics[width=\linewidth]{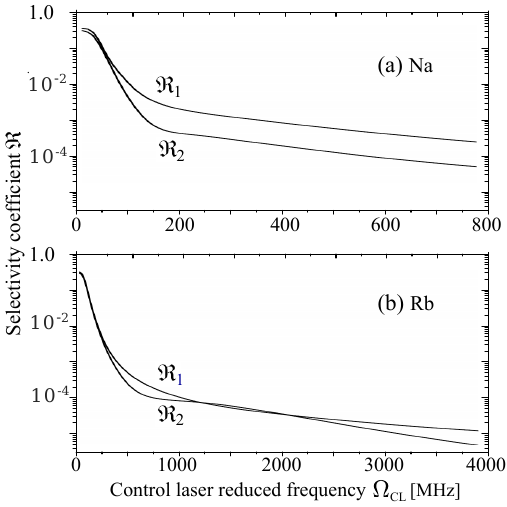}
    \caption{
    Infidelity factors $\Re _{1,2}$ (\ref{SelC}) at the resonance detunings $\Delta _P^{(Re)}$ vs the reduced frequency $\Omega _{CL}$ of the control laser for excitation
    schemes $3(5)s_{1/2}(F'')\to 3(5)p_{1/2}\!\to\! 12(14)s_{1/2}$ in the case of Na (a) and Rb (b).
    The choice of $F''\!=\!2$ results in dominant ($\Re _2 \! \ll \!1$) population of the final $F\!=\!2$ HF component while $F''\!=\!1$ corresponds to selective ($\Re _1 \! \ll\! 1$) excitation of the $F\!=\!1$ component.
    }
    \label{selectivity_RbNa}
\end{figure}
%-----------------------------Fig13_RbNa_F=1,2-end---------------

Importantly, although the state $kp_{3/2}$ as an intermediate step in the elementary ladder schemes~(\ref{mapping}) provides a rather small $\Re $, the presence of dark states $\ket{i_{0}}$ or $\ket{i_{1}}$ coupled with bright states $\ket{i}_{1}$ or $\ket{i}_{2}$ (see Tables~\ref{tab21_tab22}, \ref{tab23_tab24}) can reduce the expected efficiency of the ladders~(\ref{mapping}) as independent carriers of coherent quantum processes.
The latter circumstance can make the sequence of excitations $ks_{1/2}(F'') \!\to\! kp_{1/2}\!\to\! ns_{1/2}$ more practically attractive for the experimental implementation of selectivity due to the lack of any dark state (see \figref{control}) in the intermediate $kp_{1/2}$ step.

To circumvent the problem of strong HF population sharing between BSs $i$-states in the case of $M\!=\!1$, which prevents selectivity (see \figref{selectivity_1_2}(a) and data of \figref{selectivity_RbNa} at $\Omega _{CL}\!=\!0$), we will use the idea behind the phenomenon of electromagnetically induced transparency \cite{Shore2009,Fleischhauer2005}, namely introduce an auxiliary control laser (CL), as shown in \figref{control}.
This laser has to block the HF coupling of $\Omega _{HF}\!=\! 81.8(352.7)$MHz between BSs $\ket{i}_{F''\!=\!1,M\!=\!1}$ and $\ket{i}_{F''\!=\!2,M\!=\!1}$, providing, thus, selective excitation of the final HF components.
If one aims to dominantly populate the HF component $F\!=\!2$ (the case of \figref{selectivity_1_2}(b) and \figref{selectivity_Rb_Rb}(a)), then it is necessary to tune the control laser in accordance with \figref{control}(a); setting according to \figref{control}(b) leads to the predominant contribution of the $F\!=\!1$ component to the AT spectra in \figref{selectivity_Rb_Rb}(b).

The applied control laser has linearly polarized amplitude $\textbf{E}^{(CL)}\!=\!E^{(CL)}\textbf{e}_z$ with Gaussian spatial distribution corresponding to the switching time $\tau _{CL}\!=\! \tau _{S}\!=\!1050$ ns, therefore, its field operator $\operator{V}^{(CL)}$, the corresponding reduced Rabi $\Omega _{CL}$, and MS $\Omega ^{(CL)}_{1/2}$ frequencies are determined by Eqs.~(\ref{Field}), (\ref{rabi1}), and (\ref{RabiEf}), respectively, when replacing symbols $S \to CL$.

Figures~\ref{selectivity_1_2}-\ref{selectivity_RbNa} exhibit the results of our calculations which demonstrate significant improvement in selectivity up to the factors $\Re \sim 10^{-5}$ due to the action of the control laser even at relatively moderate MS frequencies $\Omega ^{(CL)}_{1/2} \!\simeq \! \Omega _{HF}$.
Recall (see point (ii) at the beginning of Sec.~\ref{selectivity}) that it is the values of the frequencies $\Omega ^{(CL)}_{1/2}\!\approx \! \Omega _{CL}/6$ that enter the Bloch equation ~(\ref{bloch}), thereby determining the evolution of the atomic system under study.

A discussion of how to assess the infidelity factors~(\ref{SelC}) is included in Appendix~\ref{App3}, where, in particular, the resulting Eqs.~(\ref{Re}) and (\ref{Re2}) make it possible to explain all the features of the curves $ \Re _{1,2}(\Omega _{CL})$ presented in \figref{selectivity_RbNa}.
In the region of moderate frequencies $\Omega _{CL}$, the factors $ \Re$ of deviation from selectivity associated with the HF mixing of bright states $\ket{i}_{F''=2,1}$ rapidly decrease (according to the power law $\sim \!\Omega _{CL}^{-4}$) with an increase in the control laser intensity blocking the HF interaction.
However, at large $\Omega _{CL}$, the fall in $ \Re _{1,2}(\Omega _{CL})$ slows down, turning into an inverse quadratic relationship (see Eq.~(\ref{Re2})).
It turns out that an additional channel, namely, the cascade radiative decay (RD) of excited states, contributes to the undesirable mixing of BS $\ket{i}_{F''=2,1}$.

Consider the case of the level system in \figref{control}(a), when the probe laser is tuned to excite BS $\ket{i}_{F''=2}$, which, due to the natural radiative decay into the ground state, transfers (with the rate $A_{kp}\!=\!1/\tau _{kp}$) part of its population to the HF components $\ket{g}_{F''}$ in proportion to the branching coefficients $\overline{\Pi } _{F''}\!=\!(2F''\!+\!1)/8$ \cite{Sobelman1992}.
The acquired population of the ground $\ket{g}_{F''=1}$ component is immediately, due to the control laser, transferred to the BS $\ket{i}_{F''=1}$, thus creating an additional to HF- undesirable RD-mixing between the quantum levels of the intermediate $kp_{1/2}$ state.
A similar situation is inherent in the system of levels in \figref{control}(b).

As the math in Appendix~\ref{App3} shows, RD-mixing provides an additional contribution $\Re ^{\text{(RD)}}$ to the infidelity factor (see Eq.~(\ref{Re2})), which has a quadratic drop in $\Omega _{CL}$ and therefore becomes dominant at large $\Omega _{CL}$.
Note that since the branching factor $\overline{\Pi } _{2}$ for the HF $\ket{g}_{F''=2}$ component is greater than that $\overline{\Pi } _{1}$ for $\ket{g}_{F''=1}$ by factor $5/3$, the undesirable RD-mixing is more efficient (by a factor of $25/9$) in the case of the level system in \figref{control}(b), which explains the higher position of the curve $\Re _1$ in \figref{selectivity_RbNa}(a).
Since for rubidium atoms: (i) the rate of HF-mixing is $\approx \!4.5$ times higher and (ii) the rate of RD-mixing is $\approx \tau _{5p}^{Rb}/\tau _{3p}^{Na}\!=\!1.7$ times lower compared to sodium atoms, then the frequency range with dominance of RD-mixing where $\Re _1\!>\!\Re _2$ should shift towards very high $\Omega _{CL}$-frequencies, which is clearly seen in \figref{selectivity_RbNa}(b).

%-------------------------{conclusions} begin-----------------------------

\section{Conclusions}
\label{conclusions}

Our paper investigates the formation of optically dressed states upon two-photon excitation of alkali metal atoms (see see Fig.~\ref{fig5}), focusing on the effects of constructive/destructive interference of HF atomic sublevels that can result in a modified and more restrictive two-photon selection rule for the total angular momentum.
We have described a procedure for finding a special wavefunction basis, the Morris-Shore (MS) basis, for atom interaction with linearly polarized excitation lasers.
The MS basis reduces the initially complicated multilevel excitation structure to a combination of simple mutually orthogonal three-level ladders, two-level excitation blocks, and separate isolated states (see Fig.~\ref{figSh}).
The latter are associated with dark states, while the ladder sublevels correspond to bright ones.
Two-level complexes at the second excitation step, not directly coupled to the ground $ks_{1/2}$ state by the probe laser, are related to another recently identified type of dressed states - ``Chameleon states'' \cite{Kirova2017,Cinins2021}.

Experimental observation of Autler-Townes (AT) spectra enables studying both energies and populations of dressed states, produced by a strong S-laser coupling on the second excitation step, via fluorescence response of the system (AT signal) to a weak P-laser probing the first step.
Our numerical experiments with sodium and rubidium atoms demonstrate that the AT peaks appear when the probe laser frequency is resonant with the dressed states.
Intensities of the bright peak pairs are preserved at increased Rabi frequency of the coupling field, while their resonance frequencies increasingly shift from the initial (``bare'') values at zero coupling.

Neither dark nor chameleon states contribute to AT signals, since they are decoupled from interaction with the probe laser.
However, the HF interaction operator $\operator{H}^{(hf)}$ in Eq.~(\ref{eq:Hamiltonian}) redistributes population among all the dressed states, leading to the emergence of additional, formally forbidden singlet ``dark'' components set and pairs of ``chameleon'' peaks in the AT spectra.
For high S-laser Rabi frequencies $\Omega_{S}$, the operator $\operator{H}^{(hf)}$ can be regarded as a perturbation, with its contribution to the population transfer becoming smaller and smaller as $\Omega_{S}$ increases.
As a result, these additional dark and chameleon components of the AT spectrum lose intensity, until they effectively vanish at strong coupling.
The chameleon pairs, along with the bright components, belong to the repulsive ($\pm$) branches of AT spectrum (see Fig.~\ref{4D_chameleons}) and can be further categorized into ``slow'' and ``fast'' subclasses, depending on their effective Rabi frequencies $\Omega_{3/2}^{(S)}$ (\ref{RabiEfS}) relative to the bright ones $\Omega_{1/2}^{(S)}$ \cite{Cinins2021}.
In contrast, the dark peaks tend to preserve their resonance frequencies independently of the S-laser intensity \cite{Kirova2017}.

An interesting feature of the excitation ladders $ks_{1/2}\to kp_{3/2,1/2}\to ns_{1/2}$ manifests as preservation of the HF quantum number $F$ in transitions between the initial ground $ks_{1/2}$ and the final $ns_{1/2}$ states, resulting in a specific selection rule $\Delta F\!=\!F''\!- \!F=0$ for the two-photon transitions $ks_{1/2}(F'')\to ns_{1/2}(F)$.
The reason for this phenomenon is related to the physics of the corresponding excitation processes proceeding through a series of partial two-photon paths with different intermediate HF sublevels (see Fig.~\ref{SelecEnerg_3/2} (b1)).
The destructive interference between probability amplitudes forbids the two-photon transition when $F_{f}\neq F_{g}$.
This observation opens practically important perspectives for artificially introducing custom selection rules in many-photon interactions both to manipulate quantum states and to achieve individual addressing of HF components of atomic and molecular energy levels.

In the absence of HF interaction, the control radiation of P- and S-lasers allows ideal ($100\%$) selective excitation of single three-level ladder blocks~(\ref{mapping}), i.e. excitation of selected bright $\eta$-states with a fixed $\eta$-index ($\eta =F''M$).
With them, for example, using the STIRAP technique, it is possible to carry out some basic quantum operations \cite{Vitanov2017} or, on the basis of these three $\eta $-states, to create independent cells for storing optical information by forming proper $\eta$-polaritons \cite{Vitanov2017}.
Coherent excitation of Rb Rydberg states in a three-level \cite{PhysRevLett.123.170503} or four-level \cite{Beterov2023} elementary ladder  blocks with $ F''=2, M=0$ is an important element of a three-qubit Toffoli gate implementation \cite{PhysRevLett.123.170503}. 
The data presented in Table~\ref{tab23_tab24} provides useful information for determining the optimal parameters (Rabi frequencies and laser detunings) in this type of experiments \cite{PhysRevLett.123.170503,Beterov2023} which use high-contrast Rydberg pulses.

The aforementioned selectivity, however, is violated due to the HF mixing of states at the intermediate i-step, which results in uncontrolled population leak (or information losses in applied problems) into other undesirable HF sublevels. 
Further analysis presented in subsection~\ref{ManipulationsC} and Appendix~\ref{App3} demonstrates that introduction of a third, ``blocking'' laser (control laser) enables active switching off of some off-diagonal (mixing) HF terms presented in Tables~\ref{tab21_tab22}--\ref{tab23_tab24}, and therefore allows tuning between the $|\Delta F|\leq 1$ and $|\Delta F|\equiv 0$ regimes of two-photon excitation.
As our numerical calculations for Na and Rb atoms have demonstrated, the deviation from selective population of HF components achieved in this way can be less than $0.001\%$ (see \figref{selectivity_RbNa}).
The control laser method introduced here thus opens a ``window of opportunity'' for manipulating the intra-atomic interactions, in our case eliminating the HF effects even in Rb atoms with a considerably strong HF coupling.

%============================ acknowledgments ==========================

\begin{acknowledgments}

Two of us  (A.C. and K.M.) express gratitude to the Latvian Council of Science for funding from grant No.~LZP. -2019/1-0280. 
Authors N.N.B. and I.I.R. acknowledge the support of the grant No.~23-12-00067 (https://rscf.ru/project/23-12-00067/) by the Russian Science Foundation.
We thank Klaas Bergmann for helpful discussions at the early stages of the present study.

\end{acknowledgments}

\section*{Declaration of author involvement}
\label{DKE}

Author D.K.E. declares that his participation in both the research, results of which are presented in this work, and the corresponding publication related activity was concluded by the year 2021.

%-------------Appendixes--------------

\appendix
\section{\label{App0} Semi-unitarity of field operators }

To study features of the mappings (\ref{Flip}), take two arbitrary vectors $|\alpha \rangle$, $|\tilde{\alpha }\rangle$ lying in g-subspace $\Lambda _{g}$ (or in i-subspace $\Lambda _{i}$) and consider the dot product $\langle\tilde{\beta}|\beta\rangle$ of their images $|\beta\rangle=\hat{V}^{(P(S))}|\alpha\rangle$, $|\tilde{\beta}\rangle=\hat{V}^{(P(S))}|\tilde{\alpha}\rangle$:
\begin{equation}
    \langle\tilde{\beta}|\beta\rangle=\langle \tilde{\alpha}|\hat{V}^{(P(S))\dagger}\hat{V}^{(P(S))}|\alpha\rangle;
    \label{DotProduct}
\end{equation}
\noindent The positively defined Hermitian quadratic (HQ) operators $\hat{V}^{(P)\dagger}\hat{V}^{(P)}$ and $\hat{V}^{(S)\dagger}\hat{V}^{(S)}$ act in subspaces $\Lambda_{g}$ and $\Lambda_{i}$, respectavly.
As it seen from Fig.~\ref{figbreak}, all fine-structures basis vectors in subspaces $\Lambda_{g}$ and $\Lambda_{i}$ are eigenvectors of the above HQ operators with eigenvalues equal to $|\Omega_ {1/2}^{(P)}|^2$ and $|\Omega_ {1/2}^{(S)}|^2$ ($i\! =\! 3p_{1/2}$) or $|\Omega_ {1/2,3/2}^{(S)}|^2$ ($i\! =\! 3p_{3/2}$), respectively.
The presence of only one eigenvalue implies that the corresponding operator acting in the subspace $\Lambda _{\gamma}$ is proportional to a unit operator $\hat{I}_{\gamma}$ in this subspace, i.e.
\begin{multline}
    \hat{V}^{(P)\dagger}\hat{V}^{(P)}=|\Omega_ {1/2}^{(P)}|^2 \hat{I}_g; \quad \hat{V}^{(S)\dagger}\hat{V}^{(S)}=|\Omega _{1/2}^{(S)}|^2 \hat{I}_i
    \\
    \langle\tilde{\beta'}|\beta'\rangle=|\Omega_ {1/2}^{(P)}|^2\langle \tilde{\alpha }''|\alpha ''\rangle; \quad
    \langle\tilde{\beta}|\beta\rangle=|\Omega_ {1/2}^{(S)}|^2\langle \tilde{\alpha }'|\alpha '\rangle
    \label{Orthogonalg_i}
\end{multline}
\noindent in the case of $\gamma \! =\!i\!=\!3p_{1/2}$.
The second line in Eq.~(\ref{Orthogonalg_i}) is a consequence of Eq.~(\ref{DotProduct}) and means that the field operators preserve the dot product up to a factor.

In the case of two eigenvalues $|\Omega_ {1/2,3/2}^{(S)}|^2$ ($i\! =\! 3p_{3/2}$) one can apply the following spectral decomposition for
HQ S-operator \citep{Landau1981}:
\begin{multline}
    \hat{V}^{(S)\dagger}\hat{V}^{(S)}
    =\sum_{\lambda =1/2,3/2 }|\Omega _{\lambda}^{(S)}|^2 \hat{P}_{i,\lambda}; \quad
    \hat{P}_{i, \lambda}\hat{P}_{i,\lambda} =\hat{P}_{i, \lambda}; \\
    \hat{P}_{i, 1/2}\hat{P}_{i,3/2} =0; \quad \hat{P}_{i, 1/2}+\hat{P}_{i,3/2}=\hat{I}_{i}
    \label{Prjections}
\end{multline}
\noindent Here $\hat{P}_{i,\lambda }$ denotes the projection operator to the manifold $ \Lambda _{i,\lambda} \! = \!\hat{P}_{i,\lambda} \Lambda _{i} $ of all eigenvectors corresponding to the eigenvalue $|\Omega _{\lambda}^{(S)}|^2 $, while the subspace $ \Lambda _{i} $ is the direct sum of the mutual orthogonal manifolds $ \Lambda _{i,\lambda} $: $ \Lambda _{i} \!= \!\Lambda _{i,1/2}\oplus \Lambda _{i,3/2}$ (see Fig.~\ref{figbreak} (a)). Equations~(\ref{DotProduct}) and (\ref{Prjections}) yield
\begin{equation}
    \langle\tilde{\beta}|\beta\rangle\!=\!|\Omega _{1/2}^{(S)}|^2 \langle \tilde{\alpha}'| \hat{P}_{i,1/2}|\alpha'\rangle \! +\! |\Omega _{3/2}^{(S)}|^2 \langle \tilde{\alpha}'| \hat{P}_{i,3/2}|\alpha'\rangle
    \label{DotProduct_i}
\end{equation}
\noindent So, if we take two orthogonal vectors in the subspace $\Lambda _i$, their images in the subspace $\Lambda _f$ remain orthogonal provided that both vectors belong to (1) the same manifold $\Lambda _{i, \lambda }$ or (2) to different manifolds $\Lambda _{i, 1/2 }$ and $\Lambda _{i, 3/2 }$.

Pay attention, that, as it follows from Fig.~\ref{figbreak}, the image of the ground subspace $\Lambda _{g} \equiv \Lambda _{g, 1/2}$ is equal to the manifold $\Lambda _{i, 1/2}$:
\begin{equation}
    \Lambda _{i, 1/2}=V^{(P)}\Lambda _{g}
    \label{Image}
\end{equation}

\section{\label{App1} Numerical scheme of AT spectra calculations}

For obtaining AT spectra we make use of numerical calculations of the same kind as in our previous works \cite{Kirova2017,PhysRevA.77.042511}, solving the modified Optical Bloch equations \cite{Shore2009}
\begin{align}
    \frac{d\rho}{dt}=-i\left[ \hat{H} \rho\right] + \hat{R}\rho.
    \label{bloch}
\end{align}
\noindent for atomic density matrix $\rho$ on the base of the split operator technique \cite{Kazansky,PhysRevA.77.042511}.
The total Hamiltonian $\hat{H}$ of the atom-laser system in RWA is determined by Eq.~(\ref{eq:Hamiltonian}).
The term $\hat{R}$ corresponds to relaxation processes caused by spontaneous emission and the finite laser linewidths.
The matrix representation of the Hamiltonian $\hat{H}^{a}\!+\!\hat{H}^{h}$ corresponding to isolated atomic system is a diagonal matrix in HF basis; its eigenvalues should be calculated   by accounting for the Hyperfine interaction defined in Eq.~(\ref{hfsenergy}) and Tables~\ref{tabB1},\ref{tabB2} in the next Subs.~\ref{App12}.
The field operators matrix elements (\ref{Rabi_HF}) describe a variety of stimulated transitions among the HF and Zeeman sublevels and also presented below in Eqs.~(\ref{rabi1}), (\ref{rabi2}).
In modeling rather general situations, laser intensities are chosen to be not too strong in order not to mix the HF components $F'' = 1, 2$ of the ground states  noticeably, but strong enough to mix the states of intermediate and final levels (see Fig.~\ref{fig5}).

In our calculations the Doppler broadening due to atomic velocity distribution is not taken into account.
This approximation is justified in view of our previous and forthcoming experimental
studies carried out in Doppler-free supersonic \cite{Bruvelis2012} or cold \cite{Porfido2015,Efimov2017} atomic beams, as well as in magneto-optical traps \cite{Ryabtsev2016}.

The registered AT signal $I$ belongs to the type of absorption spectra, i.e. it is proportional to the number of laser photons absorbed by one atom.
The latter, in turn, can be evaluated by summing the partial AT signals
\begin{multline}
    \label{ATsygnal}
    N_{f}\!=\! \sum_{F} N_{fF};\quad
    N_{fF}\!=\!\frac{1}{\tau _f} \int_{-\infty }^{\infty} n_{fF
    } (t) dt; \quad \\
    n_{fF} (t)\!=\!\sum_{M} \rho_{fFM,fFM}(t),
\end{multline}
\noindent each of which determines the total number of photons $ N_{fF}$ emitted by the HF component $F$ of the atomic final $f$-state with the lifetime $\tau _f$.
Here, summing over the diagonal elements of the density matrix for the $f$-state gives populations $n_{fF}$ of its HF components at the time $t$.

\subsection{\label{App12} Rabi frequencies and atomic parameters}

%-----------------------------TabB Na-beg-----------------------------
\begin{table}%[H]
    \caption{HF and radiative parameters of the coupling states in Na under consideration}
    \label{tabB1}
    \begin{tabular}{|c|c|c|c|}
    \hline
    Levels & $A_{\textit{hfs}},\,\text{MHz}$ & $B_{\textit{hfs}},\,\text{MHz}$ & $\tau,\,\text{ns}$ \\
    \hline
    $3s_{1/2}$ & 885.813 \cite{RevModPhys.49.31} & 0 & $\infty $ \\
    %\hline
    $3p_{1/2}$ & 94.44 \cite{10.1007/BF01425925} & 0 & 16.3 \cite{PhysRevLett.76.2862} \\
    %\hline
    $3p_{3/2}$ & 18.534 \cite{PhysRevA.48.1909} & 2.724 \cite{PhysRevA.48.1909} & 16.3 \cite{PhysRevLett.76.2862} \\
    %\hline
    $5s_{1/2}$ & 78.0 \cite{PhysRevLett.32.645} & 0 & 77.6 \cite{PhysRevA.30.2881} \\
    %\hline
    $4d_{3/2,5/2}$ & 0.23 \cite{Biraben1978209} & 0 & 52.4 \cite{PhysRevA.30.2881} \\
    %\hline
    $12s_{1/2}$ & 0 & 0 & 1647 \cite{PhysRevA.30.2881} \\
    \hline
    \end{tabular}
\end{table}
%-----------------------------TabB-Naend-----------------------------

%-----------------------------TabBRb-beg-----------------------------
\begin{table}%[H]
    \caption{HF and radiative parameters of the coupling states in RB \cite{article}, which are used in our simulations}
    \label{tabB2}
    \begin{tabular}{|c|c|c|c|}
    \hline
    Levels & $A_{\textit{hfs}},\,\text{MHz}$ & $B_{\textit{hfs}},\,\text{MHz}$ & $\tau,\,\text{ns}$ \\
    \hline
    $5s_{1/2}$ & 3417.3 \cite{Bize1999} & 0 & $\infty $ \\
    %\hline
    $5p_{1/2}$ & 408.3 \cite{Barwood1991} & 0 & 27.70 \cite{Volz1996} \\
    %\hline
    $5p_{3/2}$ & 84.72 \cite{article6} & 12.50\cite{article6} & 26.24 \cite{Volz1996} \\
    %\hline
    $14s_{1/2}$ & 0 & 0 & 1609 \cite{PhysRevA.30.2881} \\
    \hline
    \end{tabular}
\end{table}
%-----------------------------TabBRb-end-----------------------------

We define the parameters (Rabi frequencies, detunings) in our calculations as follows.
Each laser (with amplitudes $E_{P,S}$ and the same unit vector $\vec{e}_{z}$ of linear polarization along the quantization $z$-axis) stimulates a variety of transitions among HF and Zeeman sublevels.
It is convenient to characterize the laser induced couplings with the characteristic (reduced) Rabi frequencies $\Omega_{\Im }$ ($\Im = S,P$) \cite{PhysRevA.77.042511}
\begin{equation}
    \begin{split}
    \Omega_{P} & \equiv \Omega_{\textit{red}}^{(P)}=E^{(P)}|(3s\parallel D \parallel 3p)|; \\ \Omega_{S} & \equiv \Omega_{\textit{red}}^{(S)}=E^{(S)}|(nd,ns\parallel D \parallel 3p)|,
    \end{split}
    \label{rabi1}
\end{equation}
\noindent associated with the unresolved (with respect to the both fine and HF interactions) $g-i$ ($3s-3p$) and $i-f$ ($3p-ns$ or $3p-nd$) transitions; $(nl\parallel D \parallel n'l')$ are the corresponding reduced matrix elements \cite{Sobelman1992}.
The Rabi frequencies of individual fine ($J$) and HF ($F$) transitions $\{ lJF\to l'J'F' \}$ are defined then by the tabulated line strengths values.
Here $l$, $J$, and $F$ denote the orbital, electronic angular, and total angular momenta.
At last, the partial Rabi frequencies within the Zeeman components $M$ transitions $\{ lJFM\to l'J'F'M \}$ in the case of linear laser polarizations are evaluated via the $6j$- and $3j$-symbols as
%\cite{Sobelman1992}
\begin{multline}
    \Omega_{l'J'F'}^{(lJF)}(M) = \Omega_{\textit{red}}^{(P,S)} \frac{(nlJ\parallel D \parallel n'l'J')}{(nl\parallel D \parallel n'l')} \times \\ \left\lbrace \begin{matrix} J & F & I \\ F' & J' & 1 \end{matrix} \right\rbrace (-1)^{\Phi + F' + F - M} \times \\ \sqrt{(2F'+1)(2F+1)} \left( \begin{matrix} F & 1 & F' \\ -M & 0 & M \end{matrix} \right) ;
    \label{rabi2}
\end{multline}
\begin{multline*}
    (nlJ\parallel D \parallel n'l'J') = (-1)^{l+s+J'+1} \times \\ \sqrt{(2J'+1)(2J+1)} \left\lbrace \begin{matrix} l & J & s \\ J' & l' & 1 \end{matrix} \right\rbrace (nl\parallel D \parallel n'l'),
\end{multline*}
\noindent where $\Phi = I + J +1$ and $I = 3/2$ is the nuclear spin for both Na and Rb atoms.
The symbol $(nlJ\parallel D \parallel n'l'J')$ gives the reduced matrix element for the involved fine transition $\{ lJ\to l'J' \}$ (the electron spin is $s =1/2$).
The above expressions (\ref{rabi1}-\ref{rabi2}) unequally define the dipole interaction operator $\hat{V}^{(P,S)}$ under RWA for the given values of laser field amplitudes $E_{P,S}$, or equally, for the reduced Rabi frequencies $\Omega_{P,S}$ (\ref{rabi1}).

The atomic $\hat{H}_{a}$ and HF $\operator{H}^{(h)}$ Hamiltonian's (\ref{eq:Hamiltonian}) need for their definition information about the atomic energy structure and laser detunings.
The energy of each HF level in the system relative to the centre of mass of the respective HS manifold is given by
\begin{multline}
    \Delta \varepsilon_{\textit{hfs}}=\frac{1}{2}A_{\textit{hfs}}K + \\ B_{\textit{hfs}} \frac{3/2K(K+1)-2I(I+1)J(J+1)}{2I(2I-1)2J(2J-1)},
    \label{hfsenergy}
\end{multline}
\noindent where $K\equiv F(F+1)-I(I+1)-J(J+1)$, while $A_{\textit{hfs}}$, $B_{\textit{hfs}}$ are the magnetic dipole and electric quadrupole constants.
The detunings $\Delta_{P,S}$ of the probe and coupling fields are defined with respect to the energies $\omega^{(\textit{sp})}$ of specially selected HF sublevels of the ground ($\omega^{(\textit{sp})}_{g}=\omega_{3s(F''=2)}$), excited ($\omega^{(\textit{sp})}_{i}=\omega_{3p(F'=2)}$) and final ($\omega^{(\textit{sp})}_{f}=\omega_{5s,4d(F=2)}$) states as it is defined in the caption to \figref{fig5}.

The magnetic dipole and electric quadrupole constants, as well as the radiative lifetimes of the excited states of interest are presented in Tables~\ref{tabB1}, \ref{tabB2}.
Unless specified otherwise, the laser linewidth is assumed to be $1\,\text{MHz}$.

\section{\label{App3} Selectivity assessment under the presence of a control laser}

%-----------------------------Fig14-beg-----------------------------
\begin{figure}
    %\centering
    \includegraphics[width=\linewidth]{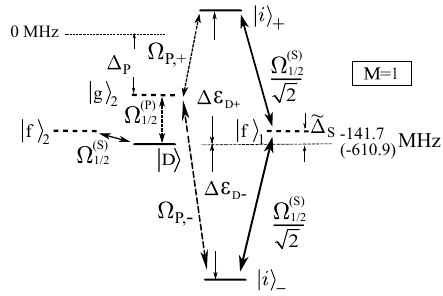}
    \caption{
    Energy levels diagrams (in MHz) for three adiabatic states $\ket{D}$, $\ket{i}_{\pm}$ in subspace $3(5)p_{1/2}(M\!=\!1)$ along with their couplings by S-laser with HF sublevels $\ket{f}_{1,2}$ of a Rydberg $ns_{1/2}$ final state and by P-laser with HF sublevel $\ket{g}_{2}$ of a ground state.
    The dashed lines depict in RWA g-sublevel $\ket{g}_{2}$ for the P-laser detuning $\Delta _P$ and the f-sublevels $\ket{f}_{1,2}$ for the S-laser detuning $\widetilde{\Delta }_S$ associated with the transition $\ket{f}_{1,2}\! \to\! \ket{i}_{2}$.
    }
    \label{control2}
\end{figure}
%-----------------------------Fig14-end-------------------------

In this appendix, we restrict ourselves to the analysis of the coefficient $\Re _2$ (\ref{SelC}) in the case of dominant excitation of the HF component $F\! = \!2$, i.e., for the working scheme of levels we will choose the scheme shown in Fig.~\ref{control}(a).
Omitted here consideration of the coefficient $\Re _1$ in the system of levels in Fig.~\ref{control}(b) with selective excitation of the $F\! = \!1$ HF component is carried out in a similar way.

The choice of the control laser (CL) detuning $\Delta_{CL}\!=\!-141.7(-610.9)\,\text{MHz}$ allows in $\Lambda$-scheme, incorporated two bright states (BS) $\ket{i}_{F''\!=\!1,2}(M\!=\!1)$ along with the RWA bare ground state $\ket{g}_{F''\!=\!1}$, to make equal the energies of its two low-lying levels shown in \figref{control}(b), frame $M\!=\!1$.
In what follows, for brevity, we will return to the notation adopted in Fig.~\ref{figSh} for BSs, i.e. write, for example, $\ket{i}_2$ instead of $\ket{i}_{F''\!=\!2}(M\!=\!1)$.

Importantly, both frequencies $ \Omega _{HF}, \Omega _{CL}$ exceed the lasers Rabi frequencies $ \Omega _{P}, \Omega _{S}$, which, therefore, can only slightly change the structure of adiabatic states formed in the considered $\Lambda $ -scheme.
The procedure for finding these adiabatic states  is well known and consists of two steps \cite{Shore2017,PhysRevA.27.906,Kirova2017}.
(i) First, two new types of dark $ \ket{D} $ and bright $ \ket{Br} $ mutually orthogonal wave vectors are constructed
\begin{multline}
    \ket{D}\!=\! \frac{\Omega _{1/2}^{(CL)}}{ \Omega _{Br}^{eff}}\ket{i}_2- \frac{\Omega _{HF}}{\Omega _{Br}^{eff}}\ket{g}_1; \quad \Omega _{Br}^{eff}\!=\!\sqrt{\Omega _{HF}^2\!+\!(\Omega _{1/2}^{(CL)})^2}
    \\
    \ket{Br}\!=\! \frac{\Omega _{HF}}{\Omega _{Br}^{eff}}\ket{i}_2\!+\! \frac{\Omega _{1/2}^{(CL)}}{ \Omega _{Br}^{eff}}\ket{g}_1,
    \label{Control_DBr}
\end{multline}
\noindent differently related to $ \ket{i}_1 $: completely decouple $D$-vector and $Br$-vector with coupling Rabi frequency $\Omega _{Br}^{eff}$.
(ii) Second, the pair of coupled diabatic vectors $\ket{Br}$, $\ket{i}_1$ is taken to find two repulsive adiabatic states $ \ket{i}_{\pm} $ defined in Eq.~(\ref{eq:DressedState}) with newly acquired energies $ \varepsilon _{i\pm} $~(\ref{eq:DressedEnergies}) due to ac Stark shifts stimulated by the coupling frequency $\Omega _{\varsigma }^{eff}\!=\!\Omega _{Br}^{eff}$ (see \figref{control2}).

Importantly, $D$-state does not change the initial diabatic energy $\varepsilon _{iD}\!=\!-141.7(-610.9)\,\text{MHz} $ corresponding to the $\ket{i}_{F''\!=\!2}(M\!=\!1)$ state, so that its energy separation $\Delta \varepsilon _{D\pm}$ from the other adiabatic $ \ket{i}_{\pm} $ states becomes $\approx \! \Omega _{Br}^{eff}$ provided that the control laser is enough intensive with $\Omega _{1/2}^{(CL) } > \Omega _{HF }$.
In this case, the diabatic vectors are evenly distributed between the adiabatic states
\begin{equation}
    \ket{i}_{\pm} \!\simeq \! \frac{1}{\sqrt{2}}\left ( \ket{i}_{1} \!\pm\!\ket{Br} \right ); \quad \Delta \varepsilon _{D\pm} \simeq \Omega _{Br}^{eff},
    \label{pm}
\end{equation}
\noindent and hence the $\ket{i}_{\pm}$ vectors-related Rabi frequency $\Omega ^{(S)}_{1/2\pm}$ of the S-laser drops by $\sqrt{2}$, i.e. $\Omega ^{(S)}_{1/2\pm}\! =\!\Omega ^{(S)}_{1/2} /\sqrt{2}$ as shown in \figref{control2}.

The wonderful property of the $D$-state is complete isolation from all other adiabatic states.
When the probe laser is tuned to $D$-state excitation, population leakage due to HF interaction is impossible, and therefore, one should expect the realization of ideal selectivity. Figures~\ref{selectivity_1_2}(b), \ref{selectivity_Rb_Rb} does show a small value of $\Re _2$-coefficient (\ref{SelC}), which, however, is not equal to zero.

Deviation from the ideal behaviour arises from the fact that the only BS $\ket{i}_2$, allowed for direct excitation by the probe laser, is represented, albeit poorly, in both $(\pm)$-vectors:
\begin{equation}
    \ket{i}_{\pm} \!\simeq \! \frac{1}{\sqrt{2}}\left ( \ket{i}_{1} \!\pm\! \frac{\Omega _{HF}}{\Omega _{Br}^{eff}}\ket{i}_2\!\pm\! \frac{\Omega ^{(CL)}_{1/2}}{\Omega _{Br}^{eff}}\ket{g}_1 \right ).
    \label{pm2}
\end{equation}
\noindent The corresponding Rabi frequencies $\Omega _{P,\pm}$ for the transitions $ \ket{g}_{2} \! \rightarrow \! \ket{i}_{\pm}$ stimulated by the probe P-laser become
\begin{equation}
    \Omega _{P,\pm}\!=\! \bra{g,F''\!=\!2M}\operator{V}^{(P)} \ket{i}_{\pm}
    \simeq \!\pm \! \frac{\Omega _{HF}}{\sqrt{2} \cdot \Omega _{Br}^{eff}}
    \Omega _{1/2}^{(P)} .
    \label{pm3}
\end{equation}
\noindent The two-photon excitation of the unwanted Rydberg HF component $\ket{f}_{1}$ from the ground HF component $\ket{g}_2$ which occurs via two virtual levels $\ket{i}_{\pm}$, is described with the following effective Rabi frequency \cite{Fleischhauer2005,Bruvelis2012}
\begin{equation}
    \Omega _{g2,\pm,f1}^{eff}\!\simeq \!2
    \frac{\Omega _{P,\pm}\Omega ^{(S)}_{1/2\pm}}{\Delta \varepsilon _{D\pm}}\!\simeq \! \frac{\Omega _{HF}\Omega ^{(S)}_{1/2}}{(\Omega _{Br}^{eff})^2}\Omega ^{(P)}_{1/2}
    \label{Effectivepm}
\end{equation}

If we take into account now that the two-photon excitation $\ket{g}_2 \! \to\!\ket{f}_2$ of  the desired Rydberg HF component $\ket{f}_{2}$ at a two-photon resonance goes via the state $\ket{D} \! \simeq \!\ket{i}_2$ (see \figref{control2} at the same energies of quantum states $\ket{f}_2 $ and $\ket{g}_2 $), the corresponding effective Rabi frequency can be roughly approximated as
\begin{equation}
    \Omega _{g2,D,f2}^{eff} \sim \frac{ \Omega ^{(P)}_{1/2} \Omega ^{(S)}_{1/2}}{\widetilde{\Delta} _S \!-\!i\Gamma _{3p}/2}
    \label{Effectivepm2}
\end{equation}
\noindent provided that the Rabi frequency $\Omega ^{(S)}_{1/2}$ does not exceed half the natural width $\Gamma _{kp}\!=\!2\pi/\tau _{kp}\!=\!9.8(5.8)$ MHz of the intimidate $kp_{1/2}$ state ($k\!=\!3(5)$).
Since the population of Rydberg components is proportional to the square of effective Rabi frequencies, the infidelity factor $\Re _2$ (18) may be estimated as
\begin{equation}
    \Re _2 \!\sim \! \frac{\Omega _{HF}^2(\widetilde{\Delta} _S^2 +\Gamma _{3p}^2/4 )}{(\Omega _{Br}^{eff})^4};\quad \Omega ^{(CL)}_{1/2}\! > \! \Omega _{HF}.
    \label{Re}
\end{equation}
\noindent At large $\Omega ^{(CL)}_{1/2}$, the effective frequency $\Omega _{Br}^{eff}$~(\ref{Control_DBr}) is reduced to $\Omega ^{(CL)}_{1/2}$, and the factor $\Re _2$ drops $\sim \Omega _{CL}^{-4}$ with increasing control laser intensity.

So far, we have ignored another process resulting in deselectivity, namely radiative decay (RD) responsible for optical pumping, which should introduce a second term into the factor $\Re _2$:
\begin{equation}
    \Re _2\!=\!\Re _2^{(\text{HF})}\!\!+\!\Re _2^{(\text{RD})}; \quad \Re _2^{(\text{HF})}\!\!\sim \! \Omega _{CL}^{-4};
    \quad \Re _2^{(\text{RD})}\!\!\sim \!\Omega _{CL}^{-2}
    \label{Re2}
\end{equation}
\noindent Below, in a brief qualitative presentation, an estimate of the asymptotic ($\Omega _{CL}\! \gg \! \Omega _{HF}$) RD contribution (term $\Re _2^{({RD})}$) to the undesirable population of the Rydberg HF component $\ket{f}_1$ will be given.

The excitation of the BS $\ket{i}_2$ is accompanied by its radiative decay into the ground $g$-state with partial population of the HF component $\ket{g}_1$, which, due to strong coupling with BS $\ket{i}_1$ (see Fig.~\ref{control}(a)), shares its population with $\ket{i}_1$ in the adiabatic states $\ket{i}_{\pm}$ (\ref{pm2}).
The latter is coupled with the component $\ket{f}_1$ by S-laser with the effective frequency $\Omega _{\pm,f1}^{eff}\!\sim \Omega _{1/2}^{(S)}$.
Since the energy defect $\simeq \!\Delta \varepsilon _{D\pm }$ between $\ket{f}_1$ and $\ket{i}_{\pm}$ determined by the control laser as $\Omega _{Br}^{eff}\!\sim \!\Omega _{CL}$ (see Eq.~(\ref{pm})) is large, the undesired $f$-state $\ket{f}_1$ can only accept the $\left (\Omega _{1/2}^{(S)}\right )^2/\Delta \varepsilon _{D\pm}^2\sim \Omega ^{-2}_{CL}$-fraction of population lost by BS $\ket{i}_2$ due RD, that yields the third term in Eq.~(\ref{Re2}).

%============================ bibliography ==========================

\bibliography{autler-townes-ii-main}

\begin{thebibliography}{49}%
\makeatletter
\providecommand \@ifxundefined [1]{%
 \@ifx{#1\undefined}
}%
\providecommand \@ifnum [1]{%
 \ifnum #1\expandafter \@firstoftwo
 \else \expandafter \@secondoftwo
 \fi
}%
\providecommand \@ifx [1]{%
 \ifx #1\expandafter \@firstoftwo
 \else \expandafter \@secondoftwo
 \fi
}%
\providecommand \natexlab [1]{#1}%
\providecommand \enquote  [1]{``#1''}%
\providecommand \bibnamefont  [1]{#1}%
\providecommand \bibfnamefont [1]{#1}%
\providecommand \citenamefont [1]{#1}%
\providecommand \href@noop [0]{\@secondoftwo}%
\providecommand \href [0]{\begingroup \@sanitize@url \@href}%
\providecommand \@href[1]{\@@startlink{#1}\@@href}%
\providecommand \@@href[1]{\endgroup#1\@@endlink}%
\providecommand \@sanitize@url [0]{\catcode `\\12\catcode `\$12\catcode
  `\&12\catcode `\#12\catcode `\^12\catcode `\_12\catcode `\%12\relax}%
\providecommand \@@startlink[1]{}%
\providecommand \@@endlink[0]{}%
\providecommand \url  [0]{\begingroup\@sanitize@url \@url }%
\providecommand \@url [1]{\endgroup\@href {#1}{\urlprefix }}%
\providecommand \urlprefix  [0]{URL }%
\providecommand \Eprint [0]{\href }%
\providecommand \doibase [0]{http://dx.doi.org/}%
\providecommand \selectlanguage [0]{\@gobble}%
\providecommand \bibinfo  [0]{\@secondoftwo}%
\providecommand \bibfield  [0]{\@secondoftwo}%
\providecommand \translation [1]{[#1]}%
\providecommand \BibitemOpen [0]{}%
\providecommand \bibitemStop [0]{}%
\providecommand \bibitemNoStop [0]{.\EOS\space}%
\providecommand \EOS [0]{\spacefactor3000\relax}%
\providecommand \BibitemShut  [1]{\csname bibitem#1\endcsname}%
\let\auto@bib@innerbib\@empty
%</preamble>
\bibitem [{\citenamefont {Ladd}\ \emph {et~al.}(2010)\citenamefont {Ladd},
  \citenamefont {Jelezko}, \citenamefont {Laflamme}, \citenamefont {Nakamura},
  \citenamefont {Monroe},\ and\ \citenamefont {O'Brien}}]{Ladd2010}%
  \BibitemOpen
  \bibfield  {author} {\bibinfo {author} {\bibfnamefont {T.~D.}\ \bibnamefont
  {Ladd}}, \bibinfo {author} {\bibfnamefont {F.}~\bibnamefont {Jelezko}},
  \bibinfo {author} {\bibfnamefont {R.}~\bibnamefont {Laflamme}}, \bibinfo
  {author} {\bibfnamefont {Y.}~\bibnamefont {Nakamura}}, \bibinfo {author}
  {\bibfnamefont {C.}~\bibnamefont {Monroe}}, \ and\ \bibinfo {author}
  {\bibfnamefont {J.~L.}\ \bibnamefont {O'Brien}},\ }\href
  {https://doi.org/10.1038/nature08812} {\bibfield  {journal} {\bibinfo
  {journal} {Nature}\ }\textbf {\bibinfo {volume} {464}},\ \bibinfo {pages}
  {45} (\bibinfo {year} {2010})}\BibitemShut {NoStop}%
\bibitem [{\citenamefont {Fowler}\ \emph {et~al.}(2012)\citenamefont {Fowler},
  \citenamefont {Mariantoni}, \citenamefont {Martinis},\ and\ \citenamefont
  {Cleland}}]{PhysRevA.86.032324}%
  \BibitemOpen
  \bibfield  {author} {\bibinfo {author} {\bibfnamefont {A.~G.}\ \bibnamefont
  {Fowler}}, \bibinfo {author} {\bibfnamefont {M.}~\bibnamefont {Mariantoni}},
  \bibinfo {author} {\bibfnamefont {J.~M.}\ \bibnamefont {Martinis}}, \ and\
  \bibinfo {author} {\bibfnamefont {A.~N.}\ \bibnamefont {Cleland}},\ }\href
  {\doibase 10.1103/PhysRevA.86.032324} {\bibfield  {journal} {\bibinfo
  {journal} {Phys. Rev. A}\ }\textbf {\bibinfo {volume} {86}},\ \bibinfo
  {pages} {032324} (\bibinfo {year} {2012})}\BibitemShut {NoStop}%
\bibitem [{\citenamefont {Saffman}(2016)}]{Saffman_2016}%
  \BibitemOpen
  \bibfield  {author} {\bibinfo {author} {\bibfnamefont {M.}~\bibnamefont
  {Saffman}},\ }\href {https://doi.org/10.1088%2F0953-4075%2F49%2F20%2F202001}
  {\bibfield  {journal} {\bibinfo  {journal} {J. Phys. B: At. Mol. Opt. Phys}\
  }\textbf {\bibinfo {volume} {49}},\ \bibinfo {pages} {202001} (\bibinfo
  {year} {2016})}\BibitemShut {NoStop}%
\bibitem [{\citenamefont {Saffman}\ \emph {et~al.}(2010)\citenamefont
  {Saffman}, \citenamefont {Walker},\ and\ \citenamefont
  {M{\o}lmer}}]{Saffman2010}%
  \BibitemOpen
  \bibfield  {author} {\bibinfo {author} {\bibfnamefont {M.}~\bibnamefont
  {Saffman}}, \bibinfo {author} {\bibfnamefont {T.~G.}\ \bibnamefont {Walker}},
  \ and\ \bibinfo {author} {\bibfnamefont {K.}~\bibnamefont {M{\o}lmer}},\
  }\href {https://doi.org/10.1103/revmodphys.82.2313} {\bibfield  {journal}
  {\bibinfo  {journal} {Rev. Mod. Phys.}\ }\textbf {\bibinfo {volume} {82}},\
  \bibinfo {pages} {2313} (\bibinfo {year} {2010})}\BibitemShut {NoStop}%
\bibitem [{\citenamefont {Ryabtsev}\ \emph {et~al.}(2016)\citenamefont
  {Ryabtsev}, \citenamefont {Beterov}, \citenamefont
  {Tret{\textquotesingle}yakov}, \citenamefont {{\`{E}}ntin},\ and\
  \citenamefont {Yakshina}}]{Ryabtsev2016}%
  \BibitemOpen
  \bibfield  {author} {\bibinfo {author} {\bibfnamefont {I.~I.}\ \bibnamefont
  {Ryabtsev}}, \bibinfo {author} {\bibfnamefont {I.~I.}\ \bibnamefont
  {Beterov}}, \bibinfo {author} {\bibfnamefont {D.~B.}\ \bibnamefont
  {Tret{\textquotesingle}yakov}}, \bibinfo {author} {\bibfnamefont {V.~M.}\
  \bibnamefont {{\`{E}}ntin}}, \ and\ \bibinfo {author} {\bibfnamefont {E.~A.}\
  \bibnamefont {Yakshina}},\ }\href
  {https://doi.org/10.3367/ufne.0186.201602k.0206} {\bibfield  {journal}
  {\bibinfo  {journal} {Phys. Usp.}\ }\textbf {\bibinfo {volume} {59}},\
  \bibinfo {pages} {196} (\bibinfo {year} {2016})}\BibitemShut {NoStop}%
\bibitem [{\citenamefont {Vitanov}\ \emph {et~al.}(2017)\citenamefont
  {Vitanov}, \citenamefont {Rangelov}, \citenamefont {Shore},\ and\
  \citenamefont {Bergmann}}]{Vitanov2017}%
  \BibitemOpen
  \bibfield  {author} {\bibinfo {author} {\bibfnamefont {N.~V.}\ \bibnamefont
  {Vitanov}}, \bibinfo {author} {\bibfnamefont {A.~A.}\ \bibnamefont
  {Rangelov}}, \bibinfo {author} {\bibfnamefont {B.~W.}\ \bibnamefont {Shore}},
  \ and\ \bibinfo {author} {\bibfnamefont {K.}~\bibnamefont {Bergmann}},\
  }\href {https://doi.org/10.1103/revmodphys.89.015006} {\bibfield  {journal}
  {\bibinfo  {journal} {Rev. Mod. Phys.}\ }\textbf {\bibinfo {volume} {89}}
  (\bibinfo {year} {2017})}\BibitemShut {NoStop}%
\bibitem [{\citenamefont {Shore}(2017)}]{Shore2017}%
  \BibitemOpen
  \bibfield  {author} {\bibinfo {author} {\bibfnamefont {B.~W.}\ \bibnamefont
  {Shore}},\ }\href {https://doi.org/10.1364/aop.9.000563} {\bibfield
  {journal} {\bibinfo  {journal} {Adv. Opt. Photonics}\ }\textbf {\bibinfo
  {volume} {9}},\ \bibinfo {pages} {563} (\bibinfo {year} {2017})}\BibitemShut
  {NoStop}%
\bibitem [{\citenamefont {Garcia-Fernandez}\ \emph {et~al.}(2005)\citenamefont
  {Garcia-Fernandez}, \citenamefont {Ekers}, \citenamefont {Yatsenko},
  \citenamefont {Vitanov},\ and\ \citenamefont
  {Bergmann}}]{PhysRevLett.95.043001}%
  \BibitemOpen
  \bibfield  {author} {\bibinfo {author} {\bibfnamefont {R.}~\bibnamefont
  {Garcia-Fernandez}}, \bibinfo {author} {\bibfnamefont {A.}~\bibnamefont
  {Ekers}}, \bibinfo {author} {\bibfnamefont {L.~P.}\ \bibnamefont {Yatsenko}},
  \bibinfo {author} {\bibfnamefont {N.~V.}\ \bibnamefont {Vitanov}}, \ and\
  \bibinfo {author} {\bibfnamefont {K.}~\bibnamefont {Bergmann}},\ }\href
  {http://link.aps.org/doi/10.1103/PhysRevLett.95.043001} {\bibfield  {journal}
  {\bibinfo  {journal} {Phys. Rev. Lett.}\ }\textbf {\bibinfo {volume} {95}},\
  \bibinfo {pages} {043001} (\bibinfo {year} {2005})}\BibitemShut {NoStop}%
\bibitem [{\citenamefont {Autler}\ and\ \citenamefont
  {Townes}(1955)}]{PhysRev.100.703}%
  \BibitemOpen
  \bibfield  {author} {\bibinfo {author} {\bibfnamefont {S.~H.}\ \bibnamefont
  {Autler}}\ and\ \bibinfo {author} {\bibfnamefont {C.~H.}\ \bibnamefont
  {Townes}},\ }\href {http://link.aps.org/doi/10.1103/PhysRev.100.703}
  {\bibfield  {journal} {\bibinfo  {journal} {Phys. Rev.}\ }\textbf {\bibinfo
  {volume} {100}},\ \bibinfo {pages} {703} (\bibinfo {year}
  {1955})}\BibitemShut {NoStop}%
\bibitem [{\citenamefont {Morris}\ and\ \citenamefont
  {Shore}(1983)}]{PhysRevA.27.906}%
  \BibitemOpen
  \bibfield  {author} {\bibinfo {author} {\bibfnamefont {J.~R.}\ \bibnamefont
  {Morris}}\ and\ \bibinfo {author} {\bibfnamefont {B.~W.}\ \bibnamefont
  {Shore}},\ }\href {http://link.aps.org/doi/10.1103/PhysRevA.27.906}
  {\bibfield  {journal} {\bibinfo  {journal} {Phys. Rev. A}\ }\textbf {\bibinfo
  {volume} {27}},\ \bibinfo {pages} {906} (\bibinfo {year} {1983})}\BibitemShut
  {NoStop}%
\bibitem [{\citenamefont {Bevilacqua}\ and\ \citenamefont
  {Arimondo}(2022)}]{Bevilacqua2022}%
  \BibitemOpen
  \bibfield  {author} {\bibinfo {author} {\bibfnamefont {G.}~\bibnamefont
  {Bevilacqua}}\ and\ \bibinfo {author} {\bibfnamefont {E.}~\bibnamefont
  {Arimondo}},\ }\href {https://doi.org/10.1088/1361-6455/ac7684} {\bibfield
  {journal} {\bibinfo  {journal} {J. Phys. B: At. Mol. Opt. Phys.}\ }\textbf
  {\bibinfo {volume} {55}},\ \bibinfo {pages} {154001} (\bibinfo {year}
  {2022})}\BibitemShut {NoStop}%
\bibitem [{\citenamefont {Kirova}\ \emph {et~al.}(2017)\citenamefont {Kirova},
  \citenamefont {Cinins}, \citenamefont {Efimov}, \citenamefont {Bruvelis},
  \citenamefont {Miculis}, \citenamefont {Bezuglov}, \citenamefont {Auzinsh},
  \citenamefont {Ryabtsev},\ and\ \citenamefont {Ekers}}]{Kirova2017}%
  \BibitemOpen
  \bibfield  {author} {\bibinfo {author} {\bibfnamefont {T.}~\bibnamefont
  {Kirova}}, \bibinfo {author} {\bibfnamefont {A.}~\bibnamefont {Cinins}},
  \bibinfo {author} {\bibfnamefont {D.~K.}\ \bibnamefont {Efimov}}, \bibinfo
  {author} {\bibfnamefont {M.}~\bibnamefont {Bruvelis}}, \bibinfo {author}
  {\bibfnamefont {K.}~\bibnamefont {Miculis}}, \bibinfo {author} {\bibfnamefont
  {N.~N.}\ \bibnamefont {Bezuglov}}, \bibinfo {author} {\bibfnamefont
  {M.}~\bibnamefont {Auzinsh}}, \bibinfo {author} {\bibfnamefont {I.~I.}\
  \bibnamefont {Ryabtsev}}, \ and\ \bibinfo {author} {\bibfnamefont
  {A.}~\bibnamefont {Ekers}},\ }\href
  {https://doi.org/10.1103/physreva.96.043421} {\bibfield  {journal} {\bibinfo
  {journal} {Phys. Rev. A}\ }\textbf {\bibinfo {volume} {96}} (\bibinfo {year}
  {2017})}\BibitemShut {NoStop}%
\bibitem [{\citenamefont {Cinins}\ \emph {et~al.}(2021)\citenamefont {Cinins},
  \citenamefont {Bruvelis}, \citenamefont {Dimitrijevi{\'{c}}}, \citenamefont
  {Sre{\'{c}}kovi{\'{c}}}, \citenamefont {Efimov}, \citenamefont {Miculis},
  \citenamefont {Bezuglov},\ and\ \citenamefont {Ekers}}]{Cinins2021}%
  \BibitemOpen
  \bibfield  {author} {\bibinfo {author} {\bibfnamefont {A.}~\bibnamefont
  {Cinins}}, \bibinfo {author} {\bibfnamefont {M.}~\bibnamefont {Bruvelis}},
  \bibinfo {author} {\bibfnamefont {M.~S.}\ \bibnamefont {Dimitrijevi{\'{c}}}},
  \bibinfo {author} {\bibfnamefont {V.~A.}\ \bibnamefont
  {Sre{\'{c}}kovi{\'{c}}}}, \bibinfo {author} {\bibfnamefont {D.~K.}\
  \bibnamefont {Efimov}}, \bibinfo {author} {\bibfnamefont {K.}~\bibnamefont
  {Miculis}}, \bibinfo {author} {\bibfnamefont {N.~N.}\ \bibnamefont
  {Bezuglov}}, \ and\ \bibinfo {author} {\bibfnamefont {A.}~\bibnamefont
  {Ekers}},\ }\href {https://doi.org/10.1002/asna.20210081} {\bibfield
  {journal} {\bibinfo  {journal} {Astron. Nachr.}\ }\textbf {\bibinfo {volume}
  {343}} (\bibinfo {year} {2021})}\BibitemShut {NoStop}%
\bibitem [{\citenamefont {Grimm}\ \emph {et~al.}(2000)\citenamefont {Grimm},
  \citenamefont {Weidem\"{u}ller},\ and\ \citenamefont
  {Ovchinnikov}}]{Grimm2000}%
  \BibitemOpen
  \bibfield  {author} {\bibinfo {author} {\bibfnamefont {R.}~\bibnamefont
  {Grimm}}, \bibinfo {author} {\bibfnamefont {M.}~\bibnamefont
  {Weidem\"{u}ller}}, \ and\ \bibinfo {author} {\bibfnamefont {Y.~B.}\
  \bibnamefont {Ovchinnikov}},\ }in\ \href
  {https://doi.org/10.1016/s1049-250x(08)60186-x} {\emph {\bibinfo {booktitle}
  {Advances In Atomic, Molecular, and Optical Physics}}}\ (\bibinfo
  {publisher} {Elsevier},\ \bibinfo {year} {2000})\ pp.\ \bibinfo {pages}
  {95--170}\BibitemShut {NoStop}%
\bibitem [{\citenamefont {Hofmann}\ \emph {et~al.}(2013)\citenamefont
  {Hofmann}, \citenamefont {G\"{u}nter}, \citenamefont {Schempp}, \citenamefont
  {M\"{u}ller}, \citenamefont {Faber}, \citenamefont {Busche}, \citenamefont
  {de~Saint-Vincent}, \citenamefont {Whitlock},\ and\ \citenamefont
  {Weidem\"{u}ller}}]{Hofmann2013}%
  \BibitemOpen
  \bibfield  {author} {\bibinfo {author} {\bibfnamefont {C.~S.}\ \bibnamefont
  {Hofmann}}, \bibinfo {author} {\bibfnamefont {G.}~\bibnamefont {G\"{u}nter}},
  \bibinfo {author} {\bibfnamefont {H.}~\bibnamefont {Schempp}}, \bibinfo
  {author} {\bibfnamefont {N.~L.~M.}\ \bibnamefont {M\"{u}ller}}, \bibinfo
  {author} {\bibfnamefont {A.}~\bibnamefont {Faber}}, \bibinfo {author}
  {\bibfnamefont {H.}~\bibnamefont {Busche}}, \bibinfo {author} {\bibfnamefont
  {M.~R.}\ \bibnamefont {de~Saint-Vincent}}, \bibinfo {author} {\bibfnamefont
  {S.}~\bibnamefont {Whitlock}}, \ and\ \bibinfo {author} {\bibfnamefont
  {M.}~\bibnamefont {Weidem\"{u}ller}},\ }\href
  {https://doi.org/10.1007/s11467-013-0396-7} {\bibfield  {journal} {\bibinfo
  {journal} {Front. Phys.}\ }\textbf {\bibinfo {volume} {9}},\ \bibinfo {pages}
  {571} (\bibinfo {year} {2013})}\BibitemShut {NoStop}%
\bibitem [{\citenamefont {Fleischhauer}\ \emph {et~al.}(2005)\citenamefont
  {Fleischhauer}, \citenamefont {Imamoglu},\ and\ \citenamefont
  {Marangos}}]{Fleischhauer2005}%
  \BibitemOpen
  \bibfield  {author} {\bibinfo {author} {\bibfnamefont {M.}~\bibnamefont
  {Fleischhauer}}, \bibinfo {author} {\bibfnamefont {A.}~\bibnamefont
  {Imamoglu}}, \ and\ \bibinfo {author} {\bibfnamefont {J.~P.}\ \bibnamefont
  {Marangos}},\ }\href {https://doi.org/10.1103/revmodphys.77.633} {\bibfield
  {journal} {\bibinfo  {journal} {Rev. Mod. Phys.}\ }\textbf {\bibinfo {volume}
  {77}},\ \bibinfo {pages} {633} (\bibinfo {year} {2005})}\BibitemShut
  {NoStop}%
\bibitem [{\citenamefont {Chu}\ and\ \citenamefont {Telnov}(2004)}]{Chu2004}%
  \BibitemOpen
  \bibfield  {author} {\bibinfo {author} {\bibfnamefont {S.-I.}\ \bibnamefont
  {Chu}}\ and\ \bibinfo {author} {\bibfnamefont {D.~A.}\ \bibnamefont
  {Telnov}},\ }\href {https://doi.org/10.1016/j.physrep.2003.10.001} {\bibfield
   {journal} {\bibinfo  {journal} {Phys. Rep.}\ }\textbf {\bibinfo {volume}
  {390}},\ \bibinfo {pages} {1} (\bibinfo {year} {2004})}\BibitemShut {NoStop}%
\bibitem [{\citenamefont {Efimov}\ \emph {et~al.}(2014)\citenamefont {Efimov},
  \citenamefont {Bezuglov}, \citenamefont {Klyucharev}, \citenamefont {Gnedin},
  \citenamefont {Miculis},\ and\ \citenamefont {Ekers}}]{efimov2014}%
  \BibitemOpen
  \bibfield  {author} {\bibinfo {author} {\bibfnamefont {D.}~\bibnamefont
  {Efimov}}, \bibinfo {author} {\bibfnamefont {N.}~\bibnamefont {Bezuglov}},
  \bibinfo {author} {\bibfnamefont {A.}~\bibnamefont {Klyucharev}}, \bibinfo
  {author} {\bibfnamefont {Y.}~\bibnamefont {Gnedin}}, \bibinfo {author}
  {\bibfnamefont {K.}~\bibnamefont {Miculis}}, \ and\ \bibinfo {author}
  {\bibfnamefont {A.}~\bibnamefont {Ekers}},\ }\href
  {http://dx.doi.org/10.1134/S0030400X1407008X} {\bibfield  {journal} {\bibinfo
   {journal} {Opt. Spectrosc.}\ }\textbf {\bibinfo {volume} {117}},\ \bibinfo
  {pages} {8} (\bibinfo {year} {2014})}\BibitemShut {NoStop}%
\bibitem [{\citenamefont {Eckardt}\ and\ \citenamefont
  {Anisimovas}(2015)}]{Eckardt2015}%
  \BibitemOpen
  \bibfield  {author} {\bibinfo {author} {\bibfnamefont {A.}~\bibnamefont
  {Eckardt}}\ and\ \bibinfo {author} {\bibfnamefont {E.}~\bibnamefont
  {Anisimovas}},\ }\href {https://doi.org/10.1088/1367-2630/17/9/093039}
  {\bibfield  {journal} {\bibinfo  {journal} {New J. Phys.}\ }\textbf {\bibinfo
  {volume} {17}},\ \bibinfo {pages} {093039} (\bibinfo {year}
  {2015})}\BibitemShut {NoStop}%
\bibitem [{\citenamefont {Shore}(2009)}]{Shore2009}%
  \BibitemOpen
  \bibfield  {author} {\bibinfo {author} {\bibfnamefont {B.}~\bibnamefont
  {Shore}},\ }\href {https://doi.org/10.1017/cbo9780511675713} {\emph {\bibinfo
  {title} {Manipulating Quantum Structures Using Laser Pulses}}}\ (\bibinfo
  {publisher} {Cambridge University Press},\ \bibinfo {year}
  {2009})\BibitemShut {NoStop}%
\bibitem [{\citenamefont {Rangelov}\ \emph {et~al.}(2006)\citenamefont
  {Rangelov}, \citenamefont {Vitanov},\ and\ \citenamefont
  {Shore}}]{Rangelov2006}%
  \BibitemOpen
  \bibfield  {author} {\bibinfo {author} {\bibfnamefont {A.~A.}\ \bibnamefont
  {Rangelov}}, \bibinfo {author} {\bibfnamefont {N.~V.}\ \bibnamefont
  {Vitanov}}, \ and\ \bibinfo {author} {\bibfnamefont {B.~W.}\ \bibnamefont
  {Shore}},\ }\href {https://doi.org/10.1103/physreva.74.053402} {\bibfield
  {journal} {\bibinfo  {journal} {Phys. Rev. A}\ }\textbf {\bibinfo {volume}
  {74}} (\bibinfo {year} {2006})}\BibitemShut {NoStop}%
\bibitem [{\citenamefont {Rousseaux}\ \emph {et~al.}(2013)\citenamefont
  {Rousseaux}, \citenamefont {Gu{\'{e}}rin},\ and\ \citenamefont
  {Vitanov}}]{Rousseaux2013}%
  \BibitemOpen
  \bibfield  {author} {\bibinfo {author} {\bibfnamefont {B.}~\bibnamefont
  {Rousseaux}}, \bibinfo {author} {\bibfnamefont {S.}~\bibnamefont
  {Gu{\'{e}}rin}}, \ and\ \bibinfo {author} {\bibfnamefont {N.~V.}\
  \bibnamefont {Vitanov}},\ }\href {https://doi.org/10.1103/physreva.87.032328}
  {\bibfield  {journal} {\bibinfo  {journal} {Phys. Rev. A}\ }\textbf {\bibinfo
  {volume} {87}} (\bibinfo {year} {2013})}\BibitemShut {NoStop}%
\bibitem [{\citenamefont {Cinins}\ \emph {et~al.}(2022)\citenamefont {Cinins},
  \citenamefont {Bruvelis},\ and\ \citenamefont {Bezuglov}}]{Cinins2022}%
  \BibitemOpen
  \bibfield  {author} {\bibinfo {author} {\bibfnamefont {A.}~\bibnamefont
  {Cinins}}, \bibinfo {author} {\bibfnamefont {M.}~\bibnamefont {Bruvelis}}, \
  and\ \bibinfo {author} {\bibfnamefont {N.~N.}\ \bibnamefont {Bezuglov}},\
  }\href {https://doi.org/10.1088/1361-6455/ac9a90} {\bibfield  {journal}
  {\bibinfo  {journal} {J. Phys. B: At. Mol. Opt. Phys.}\ }\textbf {\bibinfo
  {volume} {55}},\ \bibinfo {pages} {234003} (\bibinfo {year}
  {2022})}\BibitemShut {NoStop}%
\bibitem [{\citenamefont {Landau}\ and\ \citenamefont
  {Lifshitz}(1981)}]{Landau1981}%
  \BibitemOpen
  \bibfield  {author} {\bibinfo {author} {\bibfnamefont {L.~D.}\ \bibnamefont
  {Landau}}\ and\ \bibinfo {author} {\bibfnamefont {E.~M.}\ \bibnamefont
  {Lifshitz}},\ }\href@noop {} {\emph {\bibinfo {title} {Quantum Mechanics:
  Non-Relativistic Theory}}},\ \bibinfo {number} {ISBN: 0750635398}\ (\bibinfo
  {publisher} {Butterworth-Heinemann},\ \bibinfo {year} {1981})\BibitemShut
  {NoStop}%
\bibitem [{\citenamefont {Sobelman}(1992)}]{Sobelman1992}%
  \BibitemOpen
  \bibfield  {author} {\bibinfo {author} {\bibfnamefont {I.~I.}\ \bibnamefont
  {Sobelman}},\ }\href
  {http://www.springer.com/physics/atomic,+molecular,+optical+%26+plasma+physics/book/978-3-540-54518-7}
  {\emph {\bibinfo {title} {Atomic Spectra and Radiative Transitions}}},\
  \bibinfo {number} {ISBN: 978-3-642-76907-8}\ (\bibinfo  {publisher}
  {Springer},\ \bibinfo {year} {1992})\BibitemShut {NoStop}%
\bibitem [{\citenamefont {Delone}\ and\ \citenamefont
  {Krainov}(1999)}]{Delone_1999}%
  \BibitemOpen
  \bibfield  {author} {\bibinfo {author} {\bibfnamefont {N.~B.}\ \bibnamefont
  {Delone}}\ and\ \bibinfo {author} {\bibfnamefont {V.~P.}\ \bibnamefont
  {Krainov}},\ }\href {https://doi.org/10.1070/pu1999v042n07abeh000557}
  {\bibfield  {journal} {\bibinfo  {journal} {Phys. Usp.}\ }\textbf {\bibinfo
  {volume} {42}},\ \bibinfo {pages} {669} (\bibinfo {year} {1999})}\BibitemShut
  {NoStop}%
\bibitem [{\citenamefont {Dalibard}\ and\ \citenamefont
  {Cohen-Tannoudji}(1985)}]{Dalibard1985}%
  \BibitemOpen
  \bibfield  {author} {\bibinfo {author} {\bibfnamefont {J.}~\bibnamefont
  {Dalibard}}\ and\ \bibinfo {author} {\bibfnamefont {C.}~\bibnamefont
  {Cohen-Tannoudji}},\ }\href {https://doi.org/10.1364/josab.2.001707}
  {\bibfield  {journal} {\bibinfo  {journal} {J. Opt. Soc. Am. B: Opt. Phys.}\
  }\textbf {\bibinfo {volume} {2}},\ \bibinfo {pages} {1707} (\bibinfo {year}
  {1985})}\BibitemShut {NoStop}%
\bibitem [{\citenamefont {Bruvelis}\ \emph {et~al.}(2012)\citenamefont
  {Bruvelis}, \citenamefont {Ulmanis}, \citenamefont {Bezuglov}, \citenamefont
  {Miculis}, \citenamefont {Andreeva}, \citenamefont {Mahrov}, \citenamefont
  {Tretyakov},\ and\ \citenamefont {Ekers}}]{Bruvelis2012}%
  \BibitemOpen
  \bibfield  {author} {\bibinfo {author} {\bibfnamefont {M.}~\bibnamefont
  {Bruvelis}}, \bibinfo {author} {\bibfnamefont {J.}~\bibnamefont {Ulmanis}},
  \bibinfo {author} {\bibfnamefont {N.~N.}\ \bibnamefont {Bezuglov}}, \bibinfo
  {author} {\bibfnamefont {K.}~\bibnamefont {Miculis}}, \bibinfo {author}
  {\bibfnamefont {C.}~\bibnamefont {Andreeva}}, \bibinfo {author}
  {\bibfnamefont {B.}~\bibnamefont {Mahrov}}, \bibinfo {author} {\bibfnamefont
  {D.}~\bibnamefont {Tretyakov}}, \ and\ \bibinfo {author} {\bibfnamefont
  {A.}~\bibnamefont {Ekers}},\ }\href
  {https://doi.org/10.1103/physreva.86.012501} {\bibfield  {journal} {\bibinfo
  {journal} {Phys. Rev. A}\ }\textbf {\bibinfo {volume} {86}} (\bibinfo {year}
  {2012})}\BibitemShut {NoStop}%
\bibitem [{\citenamefont {Maxwell}\ \emph {et~al.}(2013)\citenamefont
  {Maxwell}, \citenamefont {Szwer}, \citenamefont {Paredes-Barato},
  \citenamefont {Busche}, \citenamefont {Pritchard}, \citenamefont {Gauguet},
  \citenamefont {Weatherill}, \citenamefont {Jones},\ and\ \citenamefont
  {Adams}}]{Maxwell2013}%
  \BibitemOpen
  \bibfield  {author} {\bibinfo {author} {\bibfnamefont {D.}~\bibnamefont
  {Maxwell}}, \bibinfo {author} {\bibfnamefont {D.~J.}\ \bibnamefont {Szwer}},
  \bibinfo {author} {\bibfnamefont {D.}~\bibnamefont {Paredes-Barato}},
  \bibinfo {author} {\bibfnamefont {H.}~\bibnamefont {Busche}}, \bibinfo
  {author} {\bibfnamefont {J.~D.}\ \bibnamefont {Pritchard}}, \bibinfo {author}
  {\bibfnamefont {A.}~\bibnamefont {Gauguet}}, \bibinfo {author} {\bibfnamefont
  {K.~J.}\ \bibnamefont {Weatherill}}, \bibinfo {author} {\bibfnamefont
  {M.~P.~A.}\ \bibnamefont {Jones}}, \ and\ \bibinfo {author} {\bibfnamefont
  {C.~S.}\ \bibnamefont {Adams}},\ }\href
  {https://doi.org/10.1103/physrevlett.110.103001} {\bibfield  {journal}
  {\bibinfo  {journal} {Phys. Rev. Lett.}\ }\textbf {\bibinfo {volume} {110}}
  (\bibinfo {year} {2013})}\BibitemShut {NoStop}%
\bibitem [{\citenamefont {Al-Amri}\ \emph {et~al.}(2016)\citenamefont
  {Al-Amri}, \citenamefont {El-Gomati},\ and\ \citenamefont {Zubairy}}]{2016}%
  \BibitemOpen
  \bibinfo {editor} {\bibfnamefont {M.~D.}\ \bibnamefont {Al-Amri}}, \bibinfo
  {editor} {\bibfnamefont {M.}~\bibnamefont {El-Gomati}}, \ and\ \bibinfo
  {editor} {\bibfnamefont {M.~S.}\ \bibnamefont {Zubairy}},\ eds.,\ \href
  {https://doi.org/10.1007/978-3-319-31903-2} {\emph {\bibinfo {title} {Optics
  in Our Time}}}\ (\bibinfo  {publisher} {Springer International Publishing},\
  \bibinfo {year} {2016})\BibitemShut {NoStop}%
\bibitem [{\citenamefont {L\"{o}w}\ \emph {et~al.}(2012)\citenamefont
  {L\"{o}w}, \citenamefont {Weimer}, \citenamefont {Nipper}, \citenamefont
  {Balewski}, \citenamefont {Butscher}, \citenamefont {B\"{u}chler},\ and\
  \citenamefont {Pfau}}]{Lw2012}%
  \BibitemOpen
  \bibfield  {author} {\bibinfo {author} {\bibfnamefont {R.}~\bibnamefont
  {L\"{o}w}}, \bibinfo {author} {\bibfnamefont {H.}~\bibnamefont {Weimer}},
  \bibinfo {author} {\bibfnamefont {J.}~\bibnamefont {Nipper}}, \bibinfo
  {author} {\bibfnamefont {J.~B.}\ \bibnamefont {Balewski}}, \bibinfo {author}
  {\bibfnamefont {B.}~\bibnamefont {Butscher}}, \bibinfo {author}
  {\bibfnamefont {H.~P.}\ \bibnamefont {B\"{u}chler}}, \ and\ \bibinfo {author}
  {\bibfnamefont {T.}~\bibnamefont {Pfau}},\ }\href
  {https://doi.org/10.1088/0953-4075/45/11/113001} {\bibfield  {journal}
  {\bibinfo  {journal} {J. Phys. B: At. Mol. Opt. Phys.}\ }\textbf {\bibinfo
  {volume} {45}},\ \bibinfo {pages} {113001} (\bibinfo {year}
  {2012})}\BibitemShut {NoStop}%
\bibitem [{\citenamefont {Levine}\ \emph {et~al.}(2019)\citenamefont {Levine},
  \citenamefont {Keesling}, \citenamefont {Semeghini}, \citenamefont {Omran},
  \citenamefont {Wang}, \citenamefont {Ebadi}, \citenamefont {Bernien},
  \citenamefont {Greiner}, \citenamefont {Vuleti\ifmmode~\acute{c}\else
  \'{c}\fi{}}, \citenamefont {Pichler},\ and\ \citenamefont
  {Lukin}}]{PhysRevLett.123.170503}%
  \BibitemOpen
  \bibfield  {author} {\bibinfo {author} {\bibfnamefont {H.}~\bibnamefont
  {Levine}}, \bibinfo {author} {\bibfnamefont {A.}~\bibnamefont {Keesling}},
  \bibinfo {author} {\bibfnamefont {G.}~\bibnamefont {Semeghini}}, \bibinfo
  {author} {\bibfnamefont {A.}~\bibnamefont {Omran}}, \bibinfo {author}
  {\bibfnamefont {T.~T.}\ \bibnamefont {Wang}}, \bibinfo {author}
  {\bibfnamefont {S.}~\bibnamefont {Ebadi}}, \bibinfo {author} {\bibfnamefont
  {H.}~\bibnamefont {Bernien}}, \bibinfo {author} {\bibfnamefont
  {M.}~\bibnamefont {Greiner}}, \bibinfo {author} {\bibfnamefont
  {V.}~\bibnamefont {Vuleti\ifmmode~\acute{c}\else \'{c}\fi{}}}, \bibinfo
  {author} {\bibfnamefont {H.}~\bibnamefont {Pichler}}, \ and\ \bibinfo
  {author} {\bibfnamefont {M.~D.}\ \bibnamefont {Lukin}},\ }\href {\doibase
  10.1103/PhysRevLett.123.170503} {\bibfield  {journal} {\bibinfo  {journal}
  {Phys. Rev. Lett.}\ }\textbf {\bibinfo {volume} {123}},\ \bibinfo {pages}
  {170503} (\bibinfo {year} {2019})}\BibitemShut {NoStop}%
\bibitem [{\citenamefont {Beterov}\ \emph {et~al.}(2023)\citenamefont
  {Beterov}, \citenamefont {Yakshina}, \citenamefont {Tret'yakov},
  \citenamefont {Al'yanova}, \citenamefont {Skvortsova}, \citenamefont
  {Suliman}, \citenamefont {Zagirov}, \citenamefont {Entin},\ and\
  \citenamefont {Ryabtsev}}]{Beterov2023}%
  \BibitemOpen
  \bibfield  {author} {\bibinfo {author} {\bibfnamefont {I.~I.}\ \bibnamefont
  {Beterov}}, \bibinfo {author} {\bibfnamefont {E.~A.}\ \bibnamefont
  {Yakshina}}, \bibinfo {author} {\bibfnamefont {D.~B.}\ \bibnamefont
  {Tret'yakov}}, \bibinfo {author} {\bibfnamefont {N.~V.}\ \bibnamefont
  {Al'yanova}}, \bibinfo {author} {\bibfnamefont {D.~A.}\ \bibnamefont
  {Skvortsova}}, \bibinfo {author} {\bibfnamefont {G.}~\bibnamefont {Suliman}},
  \bibinfo {author} {\bibfnamefont {T.~R.}\ \bibnamefont {Zagirov}}, \bibinfo
  {author} {\bibfnamefont {V.~M.}\ \bibnamefont {Entin}}, \ and\ \bibinfo
  {author} {\bibfnamefont {I.~I.}\ \bibnamefont {Ryabtsev}},\ }\href {\doibase
  10.1134/S1063776123080101} {\bibfield  {journal} {\bibinfo  {journal}
  {Journal of Experimental and Theoretical Physics}\ }\textbf {\bibinfo
  {volume} {137}},\ \bibinfo {pages} {246} (\bibinfo {year}
  {2023})}\BibitemShut {NoStop}%
\bibitem [{\citenamefont {Sydoryk}\ \emph {et~al.}(2008)\citenamefont
  {Sydoryk}, \citenamefont {Bezuglov}, \citenamefont {Beterov}, \citenamefont
  {Miculis}, \citenamefont {Saks}, \citenamefont {Janovs}, \citenamefont
  {Spels},\ and\ \citenamefont {Ekers}}]{PhysRevA.77.042511}%
  \BibitemOpen
  \bibfield  {author} {\bibinfo {author} {\bibfnamefont {I.}~\bibnamefont
  {Sydoryk}}, \bibinfo {author} {\bibfnamefont {N.~N.}\ \bibnamefont
  {Bezuglov}}, \bibinfo {author} {\bibfnamefont {I.~I.}\ \bibnamefont
  {Beterov}}, \bibinfo {author} {\bibfnamefont {K.}~\bibnamefont {Miculis}},
  \bibinfo {author} {\bibfnamefont {E.}~\bibnamefont {Saks}}, \bibinfo {author}
  {\bibfnamefont {A.}~\bibnamefont {Janovs}}, \bibinfo {author} {\bibfnamefont
  {P.}~\bibnamefont {Spels}}, \ and\ \bibinfo {author} {\bibfnamefont
  {A.}~\bibnamefont {Ekers}},\ }\href
  {http://link.aps.org/doi/10.1103/PhysRevA.77.042511} {\bibfield  {journal}
  {\bibinfo  {journal} {Phys. Rev. A}\ }\textbf {\bibinfo {volume} {77}},\
  \bibinfo {pages} {042511} (\bibinfo {year} {2008})}\BibitemShut {NoStop}%
\bibitem [{\citenamefont {Kazansky}\ \emph {et~al.}(2001)\citenamefont
  {Kazansky}, \citenamefont {Bezuglov}, \citenamefont {Molisch}, \citenamefont
  {Fuso},\ and\ \citenamefont {Allegrini}}]{Kazansky}%
  \BibitemOpen
  \bibfield  {author} {\bibinfo {author} {\bibfnamefont {A.~K.}\ \bibnamefont
  {Kazansky}}, \bibinfo {author} {\bibfnamefont {N.~N.}\ \bibnamefont
  {Bezuglov}}, \bibinfo {author} {\bibfnamefont {A.~F.}\ \bibnamefont
  {Molisch}}, \bibinfo {author} {\bibfnamefont {F.}~\bibnamefont {Fuso}}, \
  and\ \bibinfo {author} {\bibfnamefont {M.}~\bibnamefont {Allegrini}},\ }\href
  {http://link.aps.org/doi/10.1103/PhysRevA.64.022719} {\bibfield  {journal}
  {\bibinfo  {journal} {Phys. Rev. A}\ }\textbf {\bibinfo {volume} {64}},\
  \bibinfo {pages} {022719} (\bibinfo {year} {2001})}\BibitemShut {NoStop}%
\bibitem [{\citenamefont {Porfido}\ \emph {et~al.}(2015)\citenamefont
  {Porfido}, \citenamefont {Bezuglov}, \citenamefont {Bruvelis}, \citenamefont
  {Shayeganrad}, \citenamefont {Birindelli}, \citenamefont {Tantussi},
  \citenamefont {Guerri}, \citenamefont {Viteau}, \citenamefont {Fioretti},
  \citenamefont {Ciampini}, \citenamefont {Allegrini}, \citenamefont
  {Comparat}, \citenamefont {Arimondo}, \citenamefont {Ekers},\ and\
  \citenamefont {Fuso}}]{Porfido2015}%
  \BibitemOpen
  \bibfield  {author} {\bibinfo {author} {\bibfnamefont {N.}~\bibnamefont
  {Porfido}}, \bibinfo {author} {\bibfnamefont {N.~N.}\ \bibnamefont
  {Bezuglov}}, \bibinfo {author} {\bibfnamefont {M.}~\bibnamefont {Bruvelis}},
  \bibinfo {author} {\bibfnamefont {G.}~\bibnamefont {Shayeganrad}}, \bibinfo
  {author} {\bibfnamefont {S.}~\bibnamefont {Birindelli}}, \bibinfo {author}
  {\bibfnamefont {F.}~\bibnamefont {Tantussi}}, \bibinfo {author}
  {\bibfnamefont {I.}~\bibnamefont {Guerri}}, \bibinfo {author} {\bibfnamefont
  {M.}~\bibnamefont {Viteau}}, \bibinfo {author} {\bibfnamefont
  {A.}~\bibnamefont {Fioretti}}, \bibinfo {author} {\bibfnamefont
  {D.}~\bibnamefont {Ciampini}}, \bibinfo {author} {\bibfnamefont
  {M.}~\bibnamefont {Allegrini}}, \bibinfo {author} {\bibfnamefont
  {D.}~\bibnamefont {Comparat}}, \bibinfo {author} {\bibfnamefont
  {E.}~\bibnamefont {Arimondo}}, \bibinfo {author} {\bibfnamefont
  {A.}~\bibnamefont {Ekers}}, \ and\ \bibinfo {author} {\bibfnamefont
  {F.}~\bibnamefont {Fuso}},\ }\href
  {https://doi.org/10.1103/physreva.92.043408} {\bibfield  {journal} {\bibinfo
  {journal} {Phys. Rev. A}\ }\textbf {\bibinfo {volume} {92}} (\bibinfo {year}
  {2015})}\BibitemShut {NoStop}%
\bibitem [{\citenamefont {Efimov}\ \emph {et~al.}(2017)\citenamefont {Efimov},
  \citenamefont {Bruvelis}, \citenamefont {Bezuglov}, \citenamefont
  {Dimitrijevi{\'{c}}}, \citenamefont {Klyucharev}, \citenamefont
  {Sre{\'{c}}kovi{\'{c}}}, \citenamefont {Gnedin},\ and\ \citenamefont
  {Fuso}}]{Efimov2017}%
  \BibitemOpen
  \bibfield  {author} {\bibinfo {author} {\bibfnamefont {D.}~\bibnamefont
  {Efimov}}, \bibinfo {author} {\bibfnamefont {M.}~\bibnamefont {Bruvelis}},
  \bibinfo {author} {\bibfnamefont {N.}~\bibnamefont {Bezuglov}}, \bibinfo
  {author} {\bibfnamefont {M.}~\bibnamefont {Dimitrijevi{\'{c}}}}, \bibinfo
  {author} {\bibfnamefont {A.}~\bibnamefont {Klyucharev}}, \bibinfo {author}
  {\bibfnamefont {V.}~\bibnamefont {Sre{\'{c}}kovi{\'{c}}}}, \bibinfo {author}
  {\bibfnamefont {Y.}~\bibnamefont {Gnedin}}, \ and\ \bibinfo {author}
  {\bibfnamefont {F.}~\bibnamefont {Fuso}},\ }\href
  {https://doi.org/10.3390/atoms5040050} {\bibfield  {journal} {\bibinfo
  {journal} {Atoms}\ }\textbf {\bibinfo {volume} {5}},\ \bibinfo {pages} {50}
  (\bibinfo {year} {2017})}\BibitemShut {NoStop}%
\bibitem [{\citenamefont {Arimondo}\ \emph {et~al.}(1977)\citenamefont
  {Arimondo}, \citenamefont {Inguscio},\ and\ \citenamefont
  {Violino}}]{RevModPhys.49.31}%
  \BibitemOpen
  \bibfield  {author} {\bibinfo {author} {\bibfnamefont {E.}~\bibnamefont
  {Arimondo}}, \bibinfo {author} {\bibfnamefont {M.}~\bibnamefont {Inguscio}},
  \ and\ \bibinfo {author} {\bibfnamefont {P.}~\bibnamefont {Violino}},\ }\href
  {http://link.aps.org/doi/10.1103/RevModPhys.49.31} {\bibfield  {journal}
  {\bibinfo  {journal} {Rev. Mod. Phys.}\ }\textbf {\bibinfo {volume} {49}},\
  \bibinfo {pages} {31} (\bibinfo {year} {1977})}\BibitemShut {NoStop}%
\bibitem [{\citenamefont {Wijngaarden}\ and\ \citenamefont
  {Li}(1994)}]{10.1007/BF01425925}%
  \BibitemOpen
  \bibfield  {author} {\bibinfo {author} {\bibfnamefont {W.}~\bibnamefont
  {Wijngaarden}}\ and\ \bibinfo {author} {\bibfnamefont {J.}~\bibnamefont
  {Li}},\ }\href {http://dx.doi.org/10.1007/BF01425925} {\bibfield  {journal}
  {\bibinfo  {journal} {Z. Phys. D: At. Mol. Clusters}\ }\textbf {\bibinfo
  {volume} {32}},\ \bibinfo {pages} {67} (\bibinfo {year} {1994})}\BibitemShut
  {NoStop}%
\bibitem [{\citenamefont {Volz}\ \emph {et~al.}(1996)\citenamefont {Volz},
  \citenamefont {Majerus}, \citenamefont {Liebel}, \citenamefont {Schmitt},\
  and\ \citenamefont {Schmoranzer}}]{PhysRevLett.76.2862}%
  \BibitemOpen
  \bibfield  {author} {\bibinfo {author} {\bibfnamefont {U.}~\bibnamefont
  {Volz}}, \bibinfo {author} {\bibfnamefont {M.}~\bibnamefont {Majerus}},
  \bibinfo {author} {\bibfnamefont {H.}~\bibnamefont {Liebel}}, \bibinfo
  {author} {\bibfnamefont {A.}~\bibnamefont {Schmitt}}, \ and\ \bibinfo
  {author} {\bibfnamefont {H.}~\bibnamefont {Schmoranzer}},\ }\href
  {http://link.aps.org/doi/10.1103/PhysRevLett.76.2862} {\bibfield  {journal}
  {\bibinfo  {journal} {Phys. Rev. Lett.}\ }\textbf {\bibinfo {volume} {76}},\
  \bibinfo {pages} {2862} (\bibinfo {year} {1996})}\BibitemShut {NoStop}%
\bibitem [{\citenamefont {Yei}\ \emph {et~al.}(1993)\citenamefont {Yei},
  \citenamefont {Sieradzan},\ and\ \citenamefont {Havey}}]{PhysRevA.48.1909}%
  \BibitemOpen
  \bibfield  {author} {\bibinfo {author} {\bibfnamefont {W.}~\bibnamefont
  {Yei}}, \bibinfo {author} {\bibfnamefont {A.}~\bibnamefont {Sieradzan}}, \
  and\ \bibinfo {author} {\bibfnamefont {M.~D.}\ \bibnamefont {Havey}},\ }\href
  {http://link.aps.org/doi/10.1103/PhysRevA.48.1909} {\bibfield  {journal}
  {\bibinfo  {journal} {Phys. Rev. A}\ }\textbf {\bibinfo {volume} {48}},\
  \bibinfo {pages} {1909} (\bibinfo {year} {1993})}\BibitemShut {NoStop}%
\bibitem [{\citenamefont {Levenson}\ and\ \citenamefont
  {Bloembergen}(1974)}]{PhysRevLett.32.645}%
  \BibitemOpen
  \bibfield  {author} {\bibinfo {author} {\bibfnamefont {M.~D.}\ \bibnamefont
  {Levenson}}\ and\ \bibinfo {author} {\bibfnamefont {N.}~\bibnamefont
  {Bloembergen}},\ }\href {http://link.aps.org/doi/10.1103/PhysRevLett.32.645}
  {\bibfield  {journal} {\bibinfo  {journal} {Phys. Rev. Lett.}\ }\textbf
  {\bibinfo {volume} {32}},\ \bibinfo {pages} {645} (\bibinfo {year}
  {1974})}\BibitemShut {NoStop}%
\bibitem [{\citenamefont {Theodosiou}(1984)}]{PhysRevA.30.2881}%
  \BibitemOpen
  \bibfield  {author} {\bibinfo {author} {\bibfnamefont {C.~E.}\ \bibnamefont
  {Theodosiou}},\ }\href {http://link.aps.org/doi/10.1103/PhysRevA.30.2881}
  {\bibfield  {journal} {\bibinfo  {journal} {Phys. Rev. A}\ }\textbf {\bibinfo
  {volume} {30}},\ \bibinfo {pages} {2881} (\bibinfo {year}
  {1984})}\BibitemShut {NoStop}%
\bibitem [{\citenamefont {Biraben}\ and\ \citenamefont
  {Beroff}(1978)}]{Biraben1978209}%
  \BibitemOpen
  \bibfield  {author} {\bibinfo {author} {\bibfnamefont {F.}~\bibnamefont
  {Biraben}}\ and\ \bibinfo {author} {\bibfnamefont {K.}~\bibnamefont
  {Beroff}},\ }\href
  {http://www.sciencedirect.com/science/article/pii/0375960178901500}
  {\bibfield  {journal} {\bibinfo  {journal} {Phys. Lett. A}\ }\textbf
  {\bibinfo {volume} {65}},\ \bibinfo {pages} {209} (\bibinfo {year}
  {1978})}\BibitemShut {NoStop}%
\bibitem [{\citenamefont {Steck}(2003)}]{article}%
  \BibitemOpen
  \bibfield  {author} {\bibinfo {author} {\bibfnamefont {D.}~\bibnamefont
  {Steck}}\ }(\bibinfo {year} {2003})\BibitemShut {NoStop}%
\bibitem [{\citenamefont {Bize}\ \emph {et~al.}(1999)\citenamefont {Bize},
  \citenamefont {Sortais}, \citenamefont {Santos}, \citenamefont {Mandache},
  \citenamefont {Clairon},\ and\ \citenamefont {Salomon}}]{Bize1999}%
  \BibitemOpen
  \bibfield  {author} {\bibinfo {author} {\bibfnamefont {S.}~\bibnamefont
  {Bize}}, \bibinfo {author} {\bibfnamefont {Y.}~\bibnamefont {Sortais}},
  \bibinfo {author} {\bibfnamefont {M.~S.}\ \bibnamefont {Santos}}, \bibinfo
  {author} {\bibfnamefont {C.}~\bibnamefont {Mandache}}, \bibinfo {author}
  {\bibfnamefont {A.}~\bibnamefont {Clairon}}, \ and\ \bibinfo {author}
  {\bibfnamefont {C.}~\bibnamefont {Salomon}},\ }\href
  {https://dx.doi.org/10.1209/epl/i1999-00203-9} {\bibfield  {journal}
  {\bibinfo  {journal} {Europhys. Lett.}\ }\textbf {\bibinfo {volume} {45}},\
  \bibinfo {pages} {558} (\bibinfo {year} {1999})}\BibitemShut {NoStop}%
\bibitem [{\citenamefont {Barwood}\ \emph {et~al.}(1991)\citenamefont
  {Barwood}, \citenamefont {Gill},\ and\ \citenamefont {Rowley}}]{Barwood1991}%
  \BibitemOpen
  \bibfield  {author} {\bibinfo {author} {\bibfnamefont {G.~P.}\ \bibnamefont
  {Barwood}}, \bibinfo {author} {\bibfnamefont {P.}~\bibnamefont {Gill}}, \
  and\ \bibinfo {author} {\bibfnamefont {W.~R.~C.}\ \bibnamefont {Rowley}},\
  }\href {https://doi.org/10.1007/bf00330229} {\bibfield  {journal} {\bibinfo
  {journal} {Appl. Phys. B: Lasers Opt.}\ }\textbf {\bibinfo {volume} {53}},\
  \bibinfo {pages} {142} (\bibinfo {year} {1991})}\BibitemShut {NoStop}%
\bibitem [{\citenamefont {Volz}\ and\ \citenamefont
  {Schmoranzer}(1996)}]{Volz1996}%
  \BibitemOpen
  \bibfield  {author} {\bibinfo {author} {\bibfnamefont {U.}~\bibnamefont
  {Volz}}\ and\ \bibinfo {author} {\bibfnamefont {H.}~\bibnamefont
  {Schmoranzer}},\ }\href {https://doi.org/10.1088/0031-8949/1996/t65/007}
  {\bibfield  {journal} {\bibinfo  {journal} {Phys. Scr.}\ }\textbf {\bibinfo
  {volume} {T65}},\ \bibinfo {pages} {48} (\bibinfo {year} {1996})}\BibitemShut
  {NoStop}%
\bibitem [{\citenamefont {Ye}\ \emph {et~al.}(1996)\citenamefont {Ye},
  \citenamefont {Swartz}, \citenamefont {Jungner},\ and\ \citenamefont
  {Hall}}]{article6}%
  \BibitemOpen
  \bibfield  {author} {\bibinfo {author} {\bibfnamefont {J.}~\bibnamefont
  {Ye}}, \bibinfo {author} {\bibfnamefont {S.}~\bibnamefont {Swartz}}, \bibinfo
  {author} {\bibfnamefont {P.}~\bibnamefont {Jungner}}, \ and\ \bibinfo
  {author} {\bibfnamefont {J.}~\bibnamefont {Hall}},\ }\href@noop {} {\bibfield
   {journal} {\bibinfo  {journal} {Opt. Lett.}\ }\textbf {\bibinfo {volume}
  {21}},\ \bibinfo {pages} {1280} (\bibinfo {year} {1996})}\BibitemShut
  {NoStop}%
\end{thebibliography}%
\end{document}